%% file: confidence_pca_josse_wager_husson.tex
\documentclass[12pt]{article}

\usepackage{amssymb,dsfont}
 \usepackage{amsfonts}
\usepackage{amssymb}
\usepackage{amsmath}
\usepackage{graphicx,psfrag,epsf}
\usepackage{enumerate}
\usepackage{natbib}
\usepackage{subcaption}
\usepackage{bbm,rotating}

\newcommand{\blind}{0}

\newcommand{\bfX}{{\bf X}}

\newcommand{\bfU}{{\bf U}}
\newcommand{\bfV}{{\bf V}}

\newcommand{\bfF}{{\bf F}}
\newcommand{\bfP}{{\bf P}}

\newcommand{\bfR}{{\bf R}}
\newcommand{\bfE}{{\bf E}}

\newcommand{\bfv}{{\bf v}}

\newcommand{\bfu}{{\bf u}}
\newcommand{\bfI}{{\bf \mathbb{I}}}
\newcommand{\bfone}{{\mathds{1}}}
\newcommand{\centre}{{\left(\bfI_n-\frac{1}{n}\bfone\bfone^{\prime}\right)}}

\graphicspath{{./pic/}}

\begin{document}

\newcommand{\SPACEBIG}{1.45}
\newcommand{\SPACESMALL}{1}

\def\spacingset#1{\renewcommand{\baselinestretch}%
{#1}\small\normalsize} \spacingset{1}


\if0\blind
{
  \title{\bf Confidence Areas for Fixed-Effects PCA}
  \author{Julie Josse
	\hspace{.2cm}\\
    Applied Mathematics Department, Agrocampus Ouest, Rennes, France\\
    and \\
    Stefan  Wager \\
    Department of Statistics, Stanford University, California, USA\\
		and\\
		Fran\c cois Husson\\
		Applied Mathematics Department, Agrocampus Ouest, Rennes, France}
  \maketitle
} \fi

\if1\blind
{
  \bigskip
  \bigskip
  \bigskip
  \begin{center}
    {\LARGE\bf Confidence Areas for Fixed-Effects PCA}
\end{center}
  \medskip
} \fi

\bigskip
\begin{abstract}
PCA is often used to visualize data when the rows and the columns are both of interest. In such a setting there is a lack of inferential methods on the PCA output.
We study the asymptotic variance of a fixed-effects model for PCA, and propose several approaches to assessing the variability of PCA estimates: a method based on a parametric bootstrap, a new cell-wise jackknife, as well as a computationally cheaper approximation to the jackknife.
We visualize the confidence regions by Procrustes rotation.
Using a simulation study, we compare the proposed methods and highlight the strengths and drawbacks of each method as we vary the number of rows, the number of columns, and the strength of the relationships between variables. 
\end{abstract}

\noindent%

\spacingset{\SPACEBIG}
\section{Introduction}
\label{sec:intro}

Principal component analysis (PCA) is a well-established dimensionality reduction method used to explore and to visualize data sets. PCA is often performed on data sets that can be characterized as ``population data sets,'' where both the rows and the columns of the data matrix are fixed.
Even in this case of population data sets, we may assume that the observed data has been corrupted by random noise, and so we may want to to perform statistical inference of the PCA output. In this paper, we develop methods that let us understand the impact of noise on PCA, and to visualize this variability on PCA graphical outputs using confidence ellipses.

To give a concrete example, suppose that our dataset consists of average temperatures for the major cities in a country (rows) across the twelve months (columns); our goal is to describe the data using PCA and, e.g., to visualize which cities have similar temperature profiles and to study the correlation between the temperatures. In this example, neither the cities nor the months can be treated as random; however, it is natural to assume that there is some noise in our long-term average temperature measurements. The confidence areas developed in this paper enhance the interpretability of the PCA output. For example, they would enable us to see which cities are significantly different with respect to their climates and which cities are not.

The appropriate way of doing inference is heavily dependent on the noise model for the data matrix $\bfX$. Classical inference approaches for PCA often assume that the rows of the data matrix $\bfX$ form a random sample from a larger population.
 In this case, one of the major approaches for assessing the variability of PCA is to use non-parametric bootstrap procedures \citep{Daudin88, Holmes89, Lebart96, Timmerman07,Bada13}, which involves repeatedly re-sampling the rows of $\bfX$ with replacement.
Then, two strategies were suggested for visualizing the variability on the PCA graphical outputs: the ``partial bootstrap" and the ``total bootstrap" \citep{Greenacre84, Lebart96}. The partial bootstrap consists in projecting the $B$ replications onto the initial subspace (obtained by the PCA applied on the initial data matrix), while the total bootstrap consists in performing a PCA on each bootstrap replication. However, in order to compare the results provided by each PCA, it is necessary to minimize the ``nuisance variations" \citep{Milan95}, meaning the possible reflections or rotations between different configurations. To this end, it is possible to use Procrustes rotations \citep{Gower04} to fit the $B$ bootstrap PCA configurations toward the initial one.
Rotation problems are avoided with the partial bootstrap but \cite{Milan95} showed that this procedure underestimates the variability of the parameters.
Such a framework enables us to study the variability of the columns coordinates (or the loadings) and is well suited, for example, to situations where PCA is applied to survey data sets in which each row represents an anonymous survey participant.

We focus on an alternative noise model for the data called the ``fixed-effects" noise model where both the rows and the columns of $\bfX$ are non-random, and randomness comes only from measurement error. It is in agreement with the cities example where the rows of $\bfX$ represent specific subjects whom we want to study, and the columns of $\bfX$ represent different features of the subjects. In this case, we cannot reasonably resample the rows of $\bfX$, because the subjects we chose to study are non-random. Many examples of such data can be found in many fields. We will analyze in Section~\ref{sec:wine} a data set from sensory analysis where the rows are food products (different wines) and the columns sensory descriptors (sweetness, bitterness, etc.).

In this paper, we develop and study methods for producing confidence areas for PCA in such fixed-effects models.
We begin in Section \ref{sec:PCA} by defining a fixed-dimensional noise model for PCA. We then use this noise model to build confidence areas using various methods. In Section \ref{sec:asympt}, we define asymptotic confidence regions for PCA areas using classical theory for non-linear regression. The variability of models related to the fixed-effects PCA has been studied by \citet{denis1994} and \citet{pazman1999}, and we use their results to build these asymptotic confidence areas. We then present alternative methods based on both a parametric bootstrap (Section \ref{sec:boot}) and a cell-wise jackknife (Section \ref{sec:jackk}). As the jackknife-based method can be computationally intensive, we also propose an approximation to the jackknife based on the ``leave-out-one" lemma of \citet{Craven79}.
In order to use these confidence areas in practice, we need a way to translate them into visualizations. In  Section \ref{sec:visualisation}, we discuss Procrustes rotations that tackle problems related to the translation and rotation of different PCA configurations. 
Finally, in Section \ref{sec:Results}, we assess the coverage properties of the proposed methods using an extensive simulation study and illustrate our methods on real datasets. 

\section{Two Perspectives on Principal Component Analysis \label{sec:PCA}}

As discussed above, in order to do inference for fixed-effects PCA we first need to define a fixed-dimensional noise model. We propose such a model in Section \ref{sec:model}. Before doing so, however, we first review the geometric argument that is often used to motivate the PCA procedure.

\subsection{The Geometric Point of View}

PCA provides a low-dimensional space which minimizes the distances between points and their projections onto this subspace. 
More precisely, let $\bfX_{n \times p}$ be a (centered) matrix, and let $\|\bf A\|=\sqrt{tr(\bf A \bf A')}$ be the Hilbert-Schmidt norm.  Then, PCA corresponds to finding a matrix $\hat \bfX_{n \times p}$ of rank $S$ that provides the best approximation to the matrix $\bfX$ under least squares norm, i.e., which minimizes $\|\bfX - \hat \bfX \|^2$.
The solution is given by the singular value decomposition (SVD) of $\bfX$:
$$ \hat \bfX^{(S)} =\bfU_{n\times S} \boldsymbol{\Lambda}_{S\times S}^\frac{1}{2} \bfV_{p \times S}^{'}, $$
where $\bfU_{n\times S}$ and $\bfV_{p \times S}$ are the matrices of left and right singular vectors and $\boldsymbol{\Lambda}_{S\times S}$ is the diagonal matrix with the associated singular values. The number of components $S$ is considered here to be known.
The matrix $\bfV$ is often known as the loadings matrix, the coefficients  matrix or the principal axes matrix; the matrix $\bfF=\bfU_{n\times S} \boldsymbol{\Lambda}_{S\times S}^\frac{1}{2}$ is known as the scores matrix, the principal components matrix or as the rows coordinates matrix.

\subsection{The Model-Based Point of View}
\label{sec:model}
PCA can also be presented with a model where the data are generated as a structure corrupted by noise:
\begin{equation}
\label{eq:pca_noise}
\bfX = \tilde \bfX  + \boldsymbol{\varepsilon} 
\mbox {, }  \varepsilon_{ij} \sim \mathcal{N}(0, \sigma^2) 
\end{equation}
The structure can be considered to be random or fixed.
\citet{Caussinus86} suggested a model where the data are generated as a fixed structure having a low rank representation in $S$ dimensions corrupted by noise:
\begin{eqnarray}
\label{eq:pca_fixed}
x_{ij}  =  \sum_{s = 1}^{S} \sqrt{d_s} q_{is} r_{js}   + \varepsilon_{ij} 
\mbox {, }  \ \varepsilon_{ij} \sim \mathcal{N}(0, \sigma^2) \label{mod_acp}
\end{eqnarray}
In this model the rows have different expectations and the randomness is only due to the error term. Thus, it is more in agreement with cases where both the rows and the columns of the data matrix are of equal interest.
The maximum likelihood estimates correspond to the least squares solution, i.e., to the principal
axes and components ($\hat x_{ij}^{(S)}  =  \sum_{s = 1}^{S} \sqrt{\lambda_s} \bfu_{is} \bfv_{js}$).
This model serves as a basis to define procedures to study the variability in PCA.

Note that the noise model \eqref{eq:pca_noise} can also be paired with a random structure for $\bfX$. This leads to procedures like probabilistic PCA \citep{Tipping99a}, or the more general Factor Analysis model \citep{Barth87}.
Such random effects models are more closely in agreement with situations where the non-parametric bootstrap procedure, described in the introduction, is appropriate.

\section{Asymptotic Theory for Fixed-Effects PCA \label{sec:asympt}}

In the inferential framework associated with \eqref{eq:pca_fixed}, the number of parameters that need to be estimated grows proportionally with the size of $\bfX$. Thus, the asymptotic regime is only reached by taking the variance of the noise $\varepsilon$ down to 0. In our example presented in Section \ref{sec:wine}, the entries $x_{ij}$ are obtained by averaging the responses from multiple consumer surveys. Thus, we will naturally reach $\sigma^2 \rightarrow 0$ asymptotics as we survey more and more people.

Asymptotic confidence areas in PCA can be obtained using results originally developed for the two-factor analysis of variance problem.
We build on the work of \citet{denis1994} and \citet{pazman1999}, who studied non-linear regression models with constraints, with a focus on bi-additive models of the form
\begin{eqnarray}
y_{ij}=\mu+\alpha_i+\beta_j+\sum_{s=1}^S \gamma_{is}\delta_{js} + \varepsilon_{ij} 
 \mbox{ with}~~\varepsilon_{ij}\sim {\cal N}(0,\sigma^2). \label{biadditive} 
\end{eqnarray}
Here,  $y_{ij}$ is the response for the $(i, \, j)$ category pair,
$\mu$ is the grand mean,
$(\alpha_i)_{i=1,...,I}$ and $(\beta_j)_{j=1,...,J}$ correspond to the main effect parameters, and $$\left(\sum_{s=1}^S\gamma_{is}\delta_{js}\right)_{i=1, \, ...,\, I ; \,  j=1,\, ..., \, J}$$
models the interaction.
Such models are useful for studying the interaction between two factors when no replication is available, and have been successively applied in agronomic fields such as crop-science \citep{Gauch88, Gauch96,Cornelius96}.
\citet{denis1994, Denis96} derived the asymptotic variance of maximum likelihood estimators in the model \eqref{biadditive}.
\citet{pazman1999} then used these results to compute the bias of the parameters of model \eqref{biadditive} and showed that even if the estimators of the parameters are biased, the approximate bias of the response estimator is null.
Similar results have more recently been rediscovered by \citet{papadopoulo2000}. 
These results let us directly derive the asymptotic distribution of the PCA estimator in model \eqref{mod_acp}.  Indeed, from a computational point of view,
model \eqref{biadditive} is similar to the PCA model \eqref{mod_acp}, the main difference being that the
linear part only includes the grand mean and column main effect in PCA. 
Thus, we find that the PCA estimator
is asymptotically unbiased $\mathbb{E}(\hat x_{ij}^{(S)}) = \tilde x_{ij}$ and that the variance of the first order asymptotic approximation is
\begin{align}
&\mathbb{V}\left(\hat{x}_{ij}^{(S)}\right)=\sigma^{2}\bfP_{ij,ij}, \mbox{ where } \label{eq:va_asymp} \\
&\bfP= \left(\bfI_p\otimes \frac{1}{n}\bfone\bfone^{\prime}\right)+\left(\bfP_\bfV^{'} \otimes \centre\right) +\left(\bfI_p\otimes \bfP_\bfU\right)-\left(\bfP_\bfV^{'}\otimes \bfP_\bfU\right). \notag
\end{align}
Here, $\bfI_p$ (respectively $\bfI_n$) is the identity matrix of order $p$ (resp. of order $n$), $\bfone$ is the $n$-vector of 1's, $\otimes$ is the Kronecker product, and
$$\bfP_\bfU=\bfU( \bfU' \bfU)^{-1}\bfU' \text{ and } \bfP_\bfV= \bfV(\bfV'\bfV)^{-1} \bfV'. $$
$\bfP_\bfU$ and $\bfP_\bfV$ are the two orthogonal projection matrices involved in PCA, representing projections onto the spaces spanned by $\bfU$ and $\bfV$ respectively.
These results directly lead to  Gaussian asymptotic confidence regions. To compute these regions in practice, we draw matrices 
$$\left(\hat \bfX^{(S)^{1}}, \, ..., \, \hat \bfX^{(S)^{\star}}\right)$$
from a Gaussian distribution with expectation $\hat \bfX$ and variance given by equation \eqref{eq:va_asymp}. We will discuss in Section \ref{sec:visualisation} how to use these matrices to draw confidence areas around the row and column points in PCA graphical outputs.

The expression \eqref{eq:va_asymp} can also be derived by writing PCA as a smoothing operator.  More precisely, it can be shown \citep{Candes09, Josse11b} that PCA can be written as follows:
\begin{eqnarray}
\mbox{vec}(\hat\bfX^{(S)})&=&\bfP \mbox{vec}(\bfX),
\label{PCA_proj}
\end{eqnarray}
where $\mbox{vec}$ is the vectorization operator, i.e., $\mbox{vec}(\hat\bfX^{(S)})$ is a vector of size $np$ with the columns of $\hat\bfX^{(S)}$ stacked below each other. Here, the matrix $\bfP$ from \eqref{eq:va_asymp} acts as an orthogonal projection matrix whose elements depend on the data $\bfX$. Consequently, PCA can be seen as a ``non-linear" model where $\bfP$ represents the projection onto the tangent space of the expectation surface.
Thus, as in classical non-linear regression models, we can obtain asymptotic forms by studying linear approximations. 

Confidence areas based on linear approximations are expected to be valid as long as the non-linearity is small.
Extending prior work by \citet{Bates80} and \citet{Paz99}, \citet{Paz02} define an intrinsic measure $K_{int}$ of non-linearity for biadditive models \eqref{biadditive}:
$$K_{int}=\frac{1}{\lambda_S^{1/2}}, \text{ with } \lambda_s^{1/2}= \sum_i\gamma^2_{is}=\sum_j\delta^2_{js}. $$
 A small value of $\sigma K_{int}$ implies better accuracy of asymptotic confidence regions \citep{Bates80}. Thus, we should expect that the asymptotic confidence areas for PCA  perform best when the signal-to-noise ratio (SNR) is high. In addition, the validity of the asymptotics is all the more reliable when $n$ and $p$ are large. We assess these results in a simulation study in Section \ref{sec:Results}. 

\section{Confidence Areas for Fixed-Effects PCA \label{sec:conf}}

In this section, we propose alternatives to the asymptotic confidence areas derived above. Our methods are based both on a parametric bootstrap and the jackknife. In small samples, these non-parametric confidence areas can often out-perform their asymptotic counterpart.

\subsection{The Parametric Bootstrap \label{sec:boot}}

Bootstrap methods are often used to obtain confidence areas when the underlying model is too difficult to study analytically or before we enter the asymptotic regime. For random-effect designs where the non-parametric bootstrap procedure described in the introduction is appropriate, \citet{Timmerman07} showed that the bootstrap is more flexible and gives better confidence areas in the sense of the coverage than the corresponding asymptotic rule.

In the case of a fixed-effects model \eqref{mod_acp}, however, we cannot reasonably resample the rows of $\bfX$ as the structure of $\bfX$ is non-random. But we can still define a bootstrap procedure by regenerating residuals.
This approach is called a parametric bootstrap \citep{Efron94} as it requires an explicit model. 
It allows us to study the variability of the parameters due to the noise $\varepsilon$ in \eqref{mod_acp}, and to understand how the estimated parameters might have behaved with a different realization of the noise. This method allows us to study parameters related to both the scores and the loadings, as opposed to the non-parametric bootstrap which only lets us look at the loadings. 

We use the following parametric bootstrap algorithm:
\begin{enumerate}
\item Perform the PCA on $\bfX$ to estimate the parameters $\bfU_{n \times S}$, $\boldsymbol{\Lambda}_{S\times S}^\frac{1}{2}$, $\bfV_{p \times S}$.
\item Calculate the matrix of residuals $\hat {\boldsymbol{\varepsilon}}_{n \times p} = \bfX - \hat \bfX^{(S)}$, and estimate $\sigma^2$.
\item For $b=1,..,B$:
\begin{enumerate}
\item Draw $\varepsilon_{ij}^{b}$ from $\mathcal{N}(0,\hat \sigma^2)$ to obtain a new matrix $\boldsymbol{\varepsilon}^b$,
\item Generate a new data table: $\bfX^b = \hat \bfX^{(S)} + \boldsymbol{\varepsilon}^b$, and
\item Perform  PCA on $\bfX^b$ to obtain new estimates for the parameters $(\bfU^b, (\boldsymbol{\Lambda}^b)^{1/2}, \bfV^b)$ and a new estimator $\hat \bfX^{(S)^b}=\bfU^b(\boldsymbol{\Lambda}^b)^{\frac{1}{2}} \bfV^{b'}$.
\end{enumerate}
\end{enumerate}
This algorithm requires an estimate for the noise variance $\sigma^2$ in step 2. Since the maximum likelihood estimator is biased, \cite{Josse11b} suggested the use of
$$\hat\sigma^2=\frac{\|\bfX-\hat\bfX\|^2}{np -nS-pS+S+S^2}, $$
 which corresponds to the sum of squares of the residuals divided by the number of observed values minus the number of independent parameters estimated. 
At the end of the parametric bootstrap procedure, we have $B$ sets of estimates for the parameters $((\bfU^1, (\boldsymbol{\Lambda}^1)^{1/2}, \bfV^1), ..., (\bfU^B, (\boldsymbol{\Lambda}^{1/2})^B, \bfV^B))$ and $B$ fitted matrices $(\hat \bfX^{(S)^1} = \bfU^{1}(\boldsymbol{\Lambda^{1}})^{1/2} \bfV^{1'},..., \hat \bfX^{(S)^{B}} = \bfU^{B}(\boldsymbol{\Lambda^{B}})^{1/2} \bfV^{B'})$. We will discuss in Section \ref{sec:visualisation} how to translate them into practical confidence areas.

Since as shown in Section \ref{sec:asympt} the PCA estimator is only asymptotically unbiased ($\mathbb{E}(\hat \bfX^{(S)}) = \tilde \bfX$), we expect the coverage of the parametric bootstrap confidence areas to reach their nominal level in an asymptotic framework. 
\citet{Huet99} studied properties of confidence intervals for the parameters of non-linear models
in the asymptotic framework where the variance of the errors tends to zero, and showed that the level of confidence intervals based on the parametric residuals bootstrap are identical up to the order $o(\sigma)$ to the level of intervals calculated with first order asymptotics. Thus, in a low noise regime, we expect the parametric bootstrap procedure to give similar results to asymptotic method described in Section \ref{sec:asympt}.

\subsection{A Cell-wise Jackknife}
\label{sec:jackk} 

The jackknife is another non-parametric way to study the variability of parameters. In the context of PCA, the ``classical jackknife" procedure involves deleting each row of the data matrix $\bfX$ one at a time. \citet{Daudin89} and \citet{Besse93F} used such a procedure to define a stability index and select the number of PCA dimensions $S$.  To cut the computational costs, \citet{Daudin89} used the infinitesimal jackknife (IJ) method whereas \citet{Besse93F} used perturbation theory to define approximations. 

Removing one row at a time is analogous to the non-parametric bootstrap where the rows are considered as a random sample. Here, in the setup where the rows are not \textit{i.i.d}, we propose a new form of jackknife for PCA, which consists in removing one cell $x_{ij}$ of the data matrix at a time and estimating the PCA parameters from the incomplete data set. Writing $\hat \bfX^{(S)^{(-ij)}}$ for the PCA estimator obtained from the matrix without the cell $(ij)$, we get the pseudo-values
\begin{eqnarray}
\hat \bfX^{(S)^{(ij)}}_{jackk}=\hat \bfX +\sqrt{np} \left( \hat \bfX^{(S)^{(-ij)}} - \hat \bfX \right).
\label{eq:pseudo}
\end{eqnarray}
This procedure is repeated for each cell of the data matrix. These pseudo-values can then be transformed into confidence ellipsoids as shown in Section \ref{sec:visualisation}.

Our jackknife procedure requires a method for performing PCA with missing values. A common approach to dealing with missing values in PCA consists in minimizing the loss function over all non missing entries:
\begin{eqnarray*}
{\mathcal C} = \sum_{i=1}^{n} \sum_{j=1}^{p} w_{ij} (x_{ij}-\sum_{s=1}^{S}  \sqrt{d_s}q_{is} r_{js})^2,
\end{eqnarray*}
where $w_{ij}=0$ if $x_{ij}$ is missing and $w_{ij}=1$ otherwise.
Unlike in the complete case, there is no explicit solution to this optimization problem and it is necessary to use iterative algorithms such as alternating weighted least squares algorithms  \citep{Gabriel79}  or the expectation-maximization PCA (EM-PCA) algorithm  \citep{Kiers97,Josse12b}. This latter consists in setting the missing elements at initial values, performing the  PCA on the completed data set, filling-in the missing values using the fitted values ($\hat x_{ij}^{(S)} = \sum_{s=1}^{S}  \sqrt{\lambda_s} u_{is} v_{js}$) using a predefined  number of dimensions $S$ and repeating the procedure on the newly obtained matrix until the total change in the matrix falls below an empirically determined threshold. 

Formally, the jackknife tells us about the variability of the mean of the leave-one-out estimates. In most applications, the mean of the leave-one-out estimates is very close to the actual estimate (in the case of the sample mean, they are the same). Consequently, this specific jackknife is a competitor to both the bootstrap procedure and the method based on the asymptotic results in the sense that they estimate the same variability.
\citet[][Theorem 2]{Efr81} showed that the jackknife estimate of variance is biased upwards in expectation, suggesting that the jackknife should lead to conservative confidence areas.

\subsection{Approximating the Jackknife \label{sec:ajackk}} 

The jackknife procedure described above is computationally demanding, as it effectively requires us to run the iterative EM-PCA algorithm $n \times p$ times.
To reduce the computational effort required, we consider an approximation to the jackknife based on the ``leave-out-one'' lemma  of \citet{Craven79}. 

\citet{Josse11b} showed that the prediction error $(x_{ij}-\hat x_{ij}^{(S)^{(-ij)}})$ for a cell $(ij)$ can be approximated by
\begin{eqnarray}\label{eq:cvreg}
x_{ij}-\hat x_{ij}^{(S)^{(-ij)}} \simeq \frac{x_{ij}-\hat x_{ij}}{1-\bfP_{ij,ij}}.
\end{eqnarray}
Thus, we can define an approximation to the jackknife where instead of deleting the cell ($ij$), we replace it with the estimate $\hat x_{ij}^{(S)^{(-ij)}}$ obtained using \eqref{eq:cvreg} and then perform PCA on this altered matrix. The pseudo-values are computed as in \eqref{eq:pseudo}. The advantage of this approach is that it allows us to avoid solving missing value PCA problems.
 
Note that performing PCA on the data matrix where the cell $(ij)$ is missing (using the EM-PCA algorithm for instance) is equivalent to performing PCA on a data set where the cell $(ij)$ has been replaced with  $\hat x_{ij}^{(-ij)}$. Consequently, if the approximation \eqref{eq:cvreg} were exact, the procedure outlined here would be equivalent to the jackknife.

\section{Visualizing Confidence Areas \label{sec:visualisation}} 

All our proposed methods effectively provide a list of pseudo-realizations of the PCA estimator
\begin{equation}
\label{eq:pseudo_real}
\left\{\hat \bfX^{(S)^1} = \bfU^{1}(\boldsymbol{\Lambda^{1}})^{1/2} \bfV^{1'}, \, ..., \, \hat \bfX^{(S)^\star} = \bfU^{\star}(\boldsymbol{\Lambda^{\star}})^{1/2} \bfV^{\star'} \right\},
\end{equation}
where $\star=B$ for the parametric bootstrap and $\star=np$ for the jackknife. For example, with the bootstrap, each pseudo-realization $\hat \bfX^{(S)^b}$ corresponds to the rank-$S$ approximation obtained by performing PCA on the noised matrix $\bfX^b = \hat \bfX^{(S)} + \boldsymbol{\varepsilon}^b$. The spread of these pseudo-realizations can then give us an idea of the stability of PCA.

\spacingset{\SPACESMALL}
\begin{figure}
\begin{center}
\includegraphics[scale=0.45]{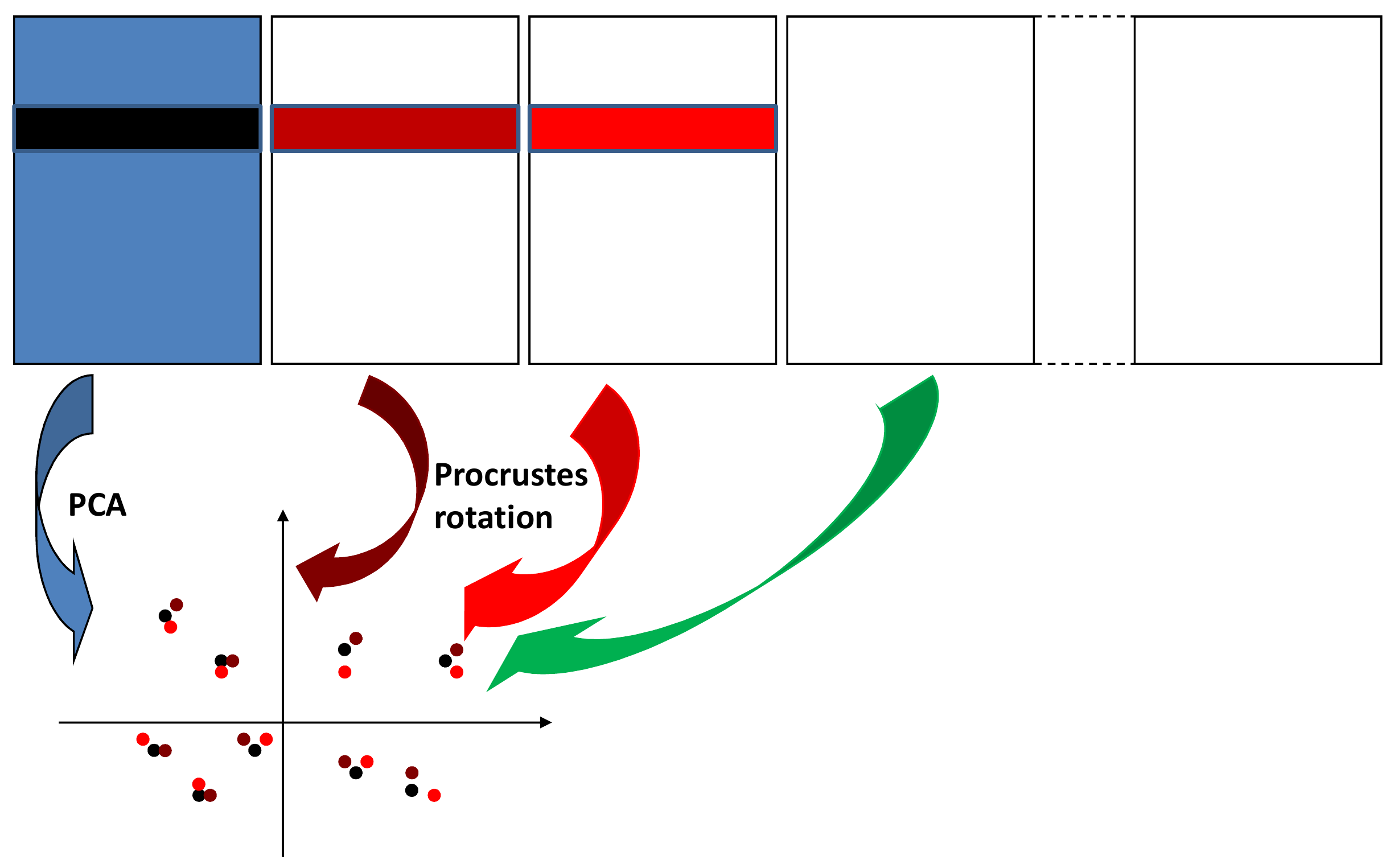}
\caption{Procrustes rotations of the rows of the PCA estimators onto the initial configuration. The first table corresponds to $\bfX$ and the other to $\hat \bfX^{(S)^1}, ..., \hat \bfX^{(S)^{\star}}$.}
\label{suppl}
\end{center}
\end{figure}
\spacingset{\SPACEBIG}

In order to make good use of the pseudo-realizations, we need to find a way to combine them into confidence areas and visualize the results. As mentioned in the introduction, this is not a trivial task since the principal components can be rotated with respect to each other from one configuration to another. At a high level, if we want to visualize confidence intervals for the rows (or columns), we need to transform $\hat \bfX^{(S)^b}$ into an approximation $\breve \bfX^{(S)^b}$ whose rows (columns) are contained in the row- (column-) span of $\hat\bfX^{(S)}$. Once we have made this transformation, we can generate confidence ellipses by looking at the variance of the $\breve \bfX^{(S)^b}$.

We address the problem of transforming $\hat \bfX^{(S)^b}$ into the appropriate space of $\hat \bfX^{(S)}$ using Procrustes rotation.  \citep{Gower04,Krza00,Schoe66}, which is often used to compare the results of different PCA analyses \citep{Lebart07}. Recall that all our matrices are already centered, so we do not need to worry about translation. Suppose that we want to study the rows of $\bfX$. Then, we define $\breve \bfX^{(S)^b}$ as
\begin{equation}
\label{eq:proc_rot}
\breve \bfX^{(S)^b} =   \hat \bfX^{(S)^b} \, \bfR^b \text{ where } \bfR^b = \operatorname{argmin}_{\bfR} \left\{ \left\lVert \hat \bfX^{(S)} -  \hat \bfX^{(S)^b } \, \bfR\right\rVert^2  : \bfR^{\prime} \bfR = \mathbb{I}_p \right\},
\end{equation}
i.e., $\breve \bfX^{(S)^b}$ is the best rotation of $\hat \bfX^{(S)^b}$ towards $\hat \bfX^{(S)}$. The well-know solution to \eqref{eq:proc_rot} is to set $\bfR^b= \bfE_{l} \bfE^{\prime}_{r}$, where $\bfE_{l}$ and $\bfE_{r}$ represent the left and right singular vectors of the matrix $(\hat\bfX^{(S)^{b}})^{\prime} \hat\bfX^{(S)}$. As desired, the rows of $\breve \bfX^{(S)^b}$ are contained in the row-span of $\hat\bfX^{(S)}$ by construction.

Finally, we note that our procedure relies heavily on our ability to generate pseudo-realizations of the form \eqref{eq:pseudo_real}. Thus, it cannot be applied to the non-parametric bootstrap (which involves resampling rows) or to the classical jackknife (which deletes one row at a time). Both of these approaches modify the structure of the rows of $\bfX$ for each bootstrap replication, and so do not produce quantities of the form \eqref{eq:pseudo_real}.

\section{Results \label{sec:Results}}

To assess our proposed methods, we first run a comparative simulation study below, and then apply the methods to a wine dataset in Section \ref{sec:wine} and a decathlon dataset in Section \ref{sec:decathlon}.

\subsection{Simulations  \label{sec:simu} }

\spacingset{\SPACESMALL}
\input{./table1.tex}
\spacingset{\SPACEBIG}

For our simulation study, we generated data according to the model~\eqref{mod_acp} with $S=2$ dimensions. 
More precisely, $\tilde{\bfX}$ is generated as follows:
\begin{enumerate}
\item A SVD is performed on an $n \times S$ matrix generated from a standard multivariate normal distribution. The left singular vectors provide $S$ empirically orthonormal vectors.
\item Each vector $s: \, 1, \, ..., \, S$ is replicated to obtain $p$ variables such that the ratio between the eigenvalues is fixed to a pre-determined value $d_1/d_2$. For instance, if $p=50$ and $d_1/d_2 = 4$, the first vector is replicated 40 times and the second vector is replicated 10 times.
\item Then, to generate the data matrix $\bfX$, we add an isotropic Gaussian noise to the structure. The scale of the Gaussian noise is determined by the target signal-to-noise (SNR) ratio.
\end{enumerate}
Here, a high SNR implies that the columns of $\bfX$ are highly correlated, whereas a low SNR implies that the data is very noisy. For each combination of the parameters, we generated 50 data sets.
Having generated data matrices $\bfX$, we ran all our proposed methods to generate confidence ellipses for the PCA representation as discussed in Section \ref{sec:visualisation}, and then evaluated the coverage of the ellipses. To do so, we counted the number of times that the row points of the true configuration $\tilde \bfX$ are inside the ellipses. Results of the simulation study are gathered in Table~\ref{tab:coverage}.

As our simulation study makes clear, the proposed methods have very different strengths and weaknesses. The first two methods, namely the asymptotic method and the parametric bootstrap, both perform similarly although the bootstrap is somewhat better.
Both methods provide accurate confidence areas close to the nominal level when the SNR is high and when $n$ is large, but can break down in low SNR settings. This is not surprising, as the asymptotic method was justified by a linear approximation which is valid in the low noise limit, and the parametric bootstrap should behave similarly to the asymptotic method when both are valid.

Conversely, the jackknife and its approximation behave well in low SNR settings, but can struggle when $p$ is small or when there is not much noise. In general, the exact jackknife is more accurate, but can sometimes be prohibitively expensive computationally.

\spacingset{\SPACESMALL}
\begin{figure}
\centering
\begin{tabular}{cc}
\includegraphics[width = 0.4\columnwidth]{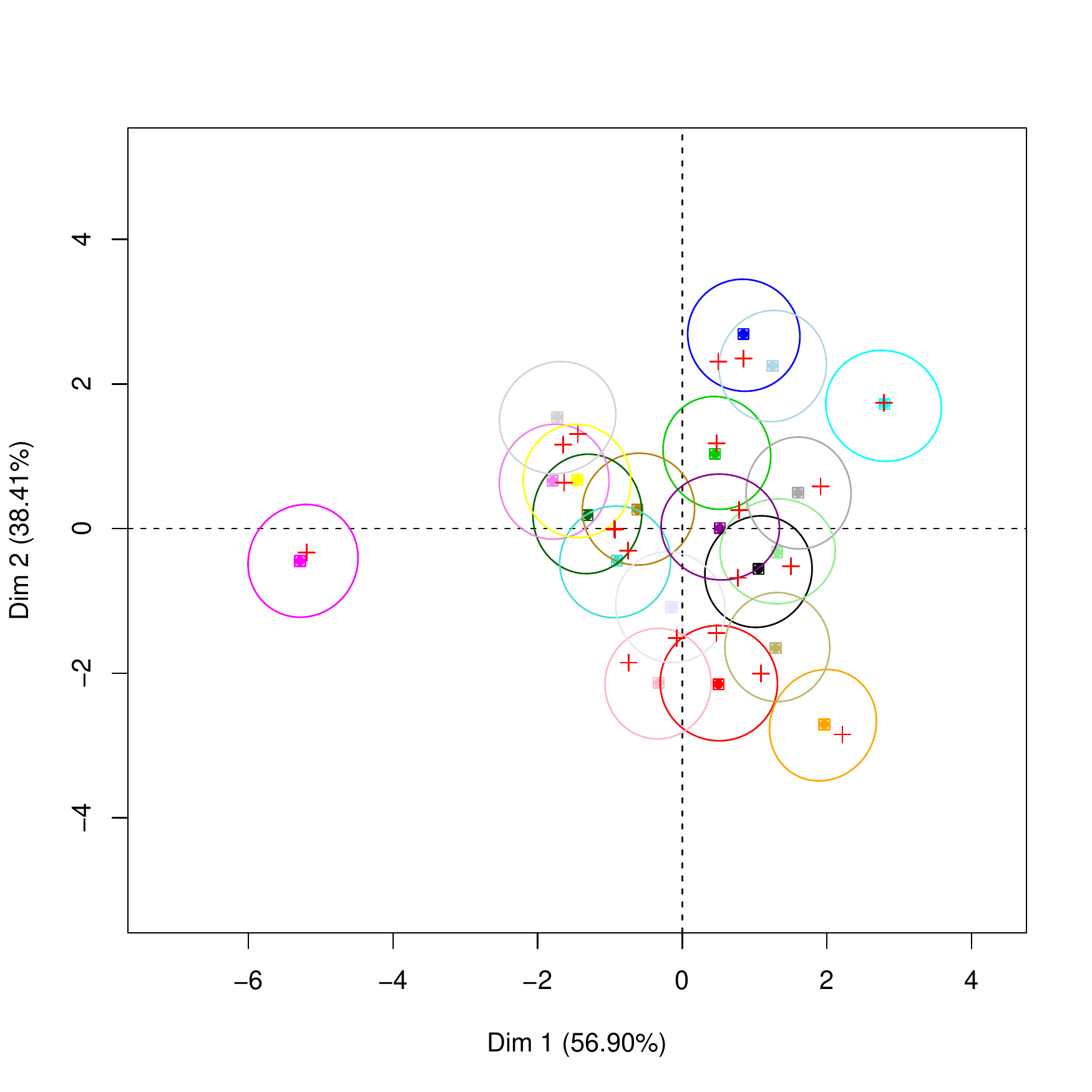} &
\includegraphics[width = 0.4\columnwidth]{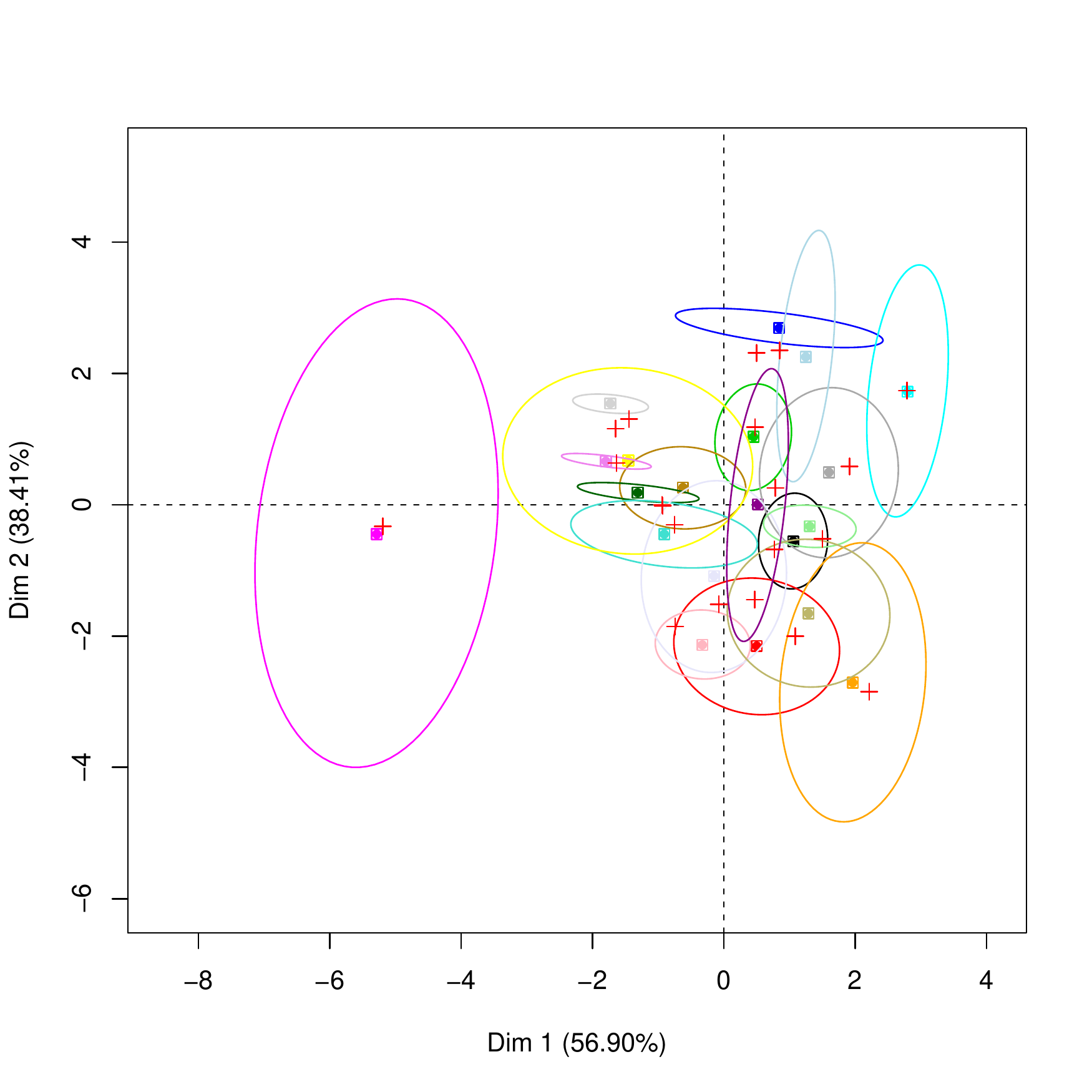} \\
Parametric Bootstrap & Jackknife \\
\end{tabular}
\caption{Simulation with $n=20$, $p=5$, $SNR=4$, $d1/d2 = 1$. The ellipses represent confidence areas for the two-dimensional representations of the row points produced by PCA. The row points of the true configuration  $\tilde \bfX$ are represented with red crosses.}
\label{fig:simu_A}
\end{figure} 
\spacingset{\SPACEBIG}
 
To gain insight into these results we focus on specific cases.  First, we consider a small matrix ($n=20$, $p=5$, SNR=4, $d_1/d_2 = 1$) for which the jackknife had a lot of trouble. As seen in Figure \ref{fig:simu_A}, the confidence ellipsoids produced by the jackknife are highly variable: some of them are much too big, whereas others are much too small. This variability then makes the jackknife have poor average coverage. Meanwhile, the bootstrap here achieves stable variance estimates and good coverage.

The reason for this instability appears to be that the jackknife ellipses are each dominated by very few pseudo-realizations. It seems that the ellipse for each row point is governed by only $p = 5$ pseudo-realizations obtained by leaving out cells from that row. Moreover, even the few pseudo-realizations we have appear to all be fairly correlated. This may be an artifact of our data generation procedure, where many columns are almost identical in the low noise setup. This correlation between columns could also explain why the jackknife deals rather poorly with high SNR in our simulation, even when $p$ is larger than 5.

\spacingset{\SPACESMALL}
\begin{figure}
\centering
\begin{tabular}{cc}
\includegraphics[width = 0.4\columnwidth]{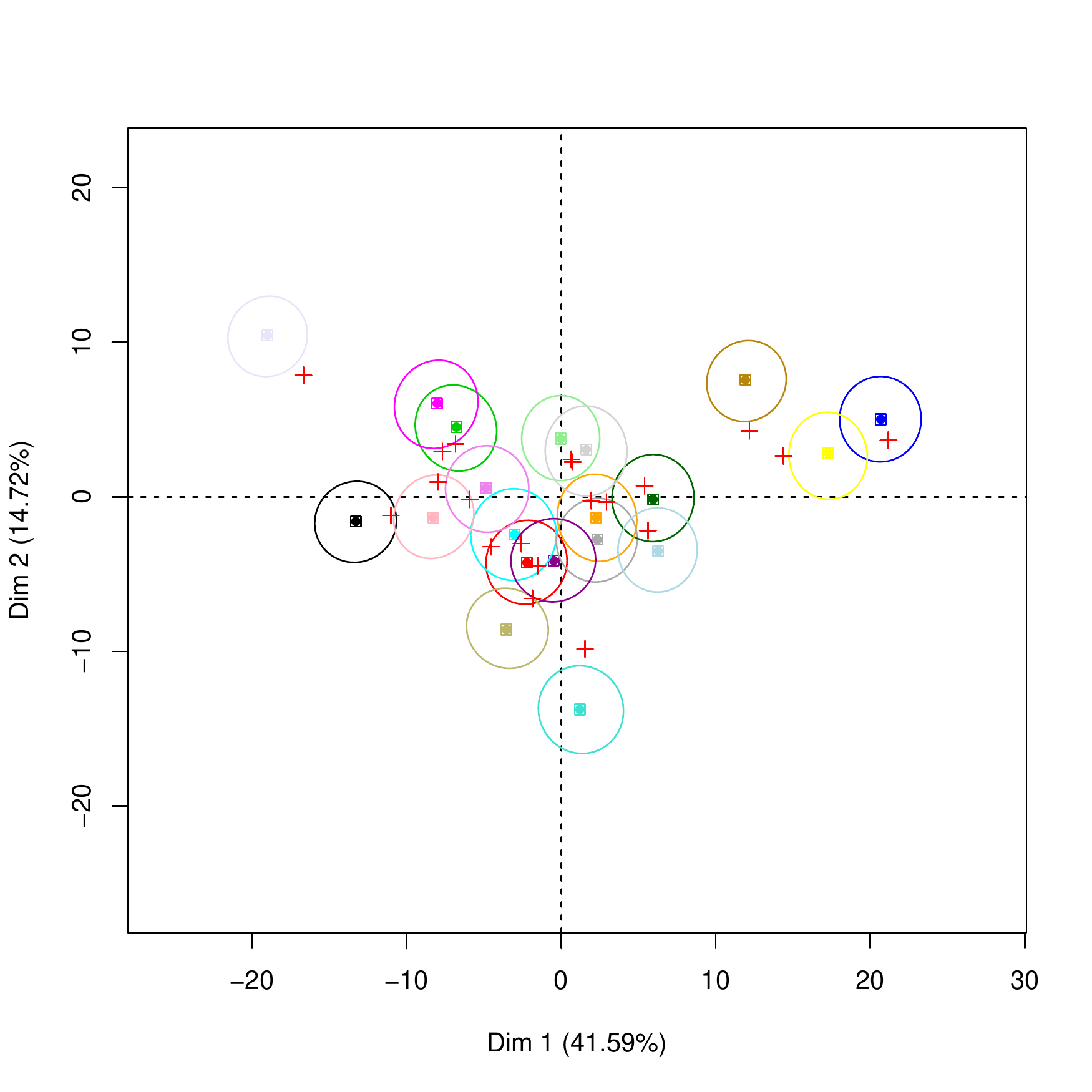} &
\includegraphics[width = 0.4\columnwidth]{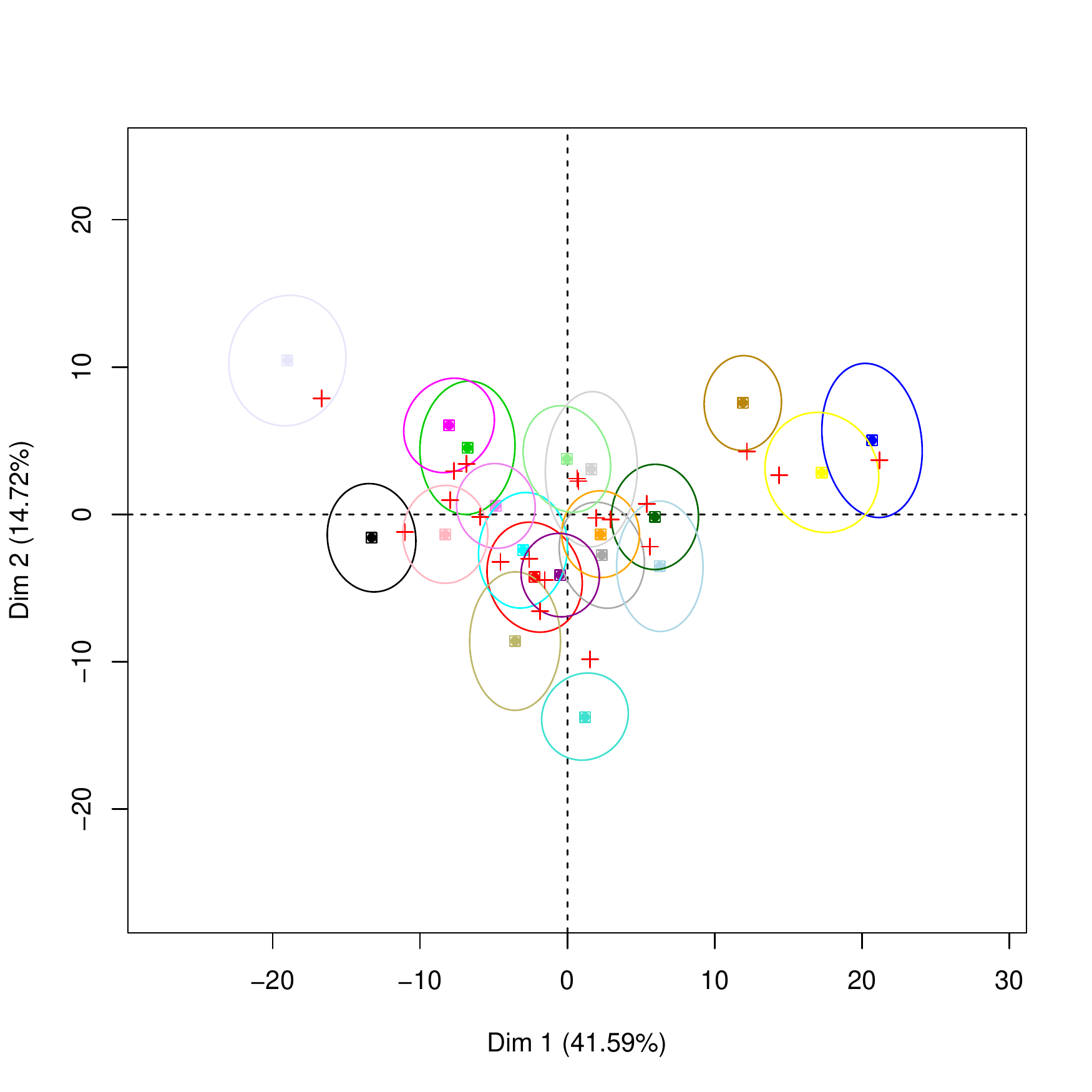} \\
Parametric Bootstrap & Jackknife \\
\end{tabular}
\caption{One simulation with $n=20$, $p=100$, $SNR=1$, $d1/d2 = 4$. The ellipses represent confidence areas for the two-dimensional representations of the row points produced by PCA. The row points of the true configuration  $\tilde \bfX$ are represented with red crosses.}
\label{fig:simu_B}
\end{figure}
\spacingset{\SPACEBIG}

Figure \ref{fig:simu_B} illustrates a lower signal case ($n=20$, $p=100$, SNR=1, $d_1/d_2 = 4$), for which the bootstrap has more trouble. We see that, here, the bootstrap appears to be over-fitting the data and under-estimating the size of the noise;  the bootstrap ellipses are too small and often do not contain the true values. Meanwhile, the jackknife appears to be performing reasonably well here. In repeated simulations (not shown in Table \ref{tab:coverage}), the methods achieved average coverage proportions of 0.79 for the asymptotic method, 0.83 for the bootstrap, 0.91 for the jackknife and 0.89 for the approximate jackknife.

\subsection{Wine Data}
\label{sec:wine}

\spacingset{\SPACESMALL}
\begin{figure}
\begin{center}
\begin{subfigure}{0.48\textwidth}
\centering
\includegraphics[width=\textwidth]{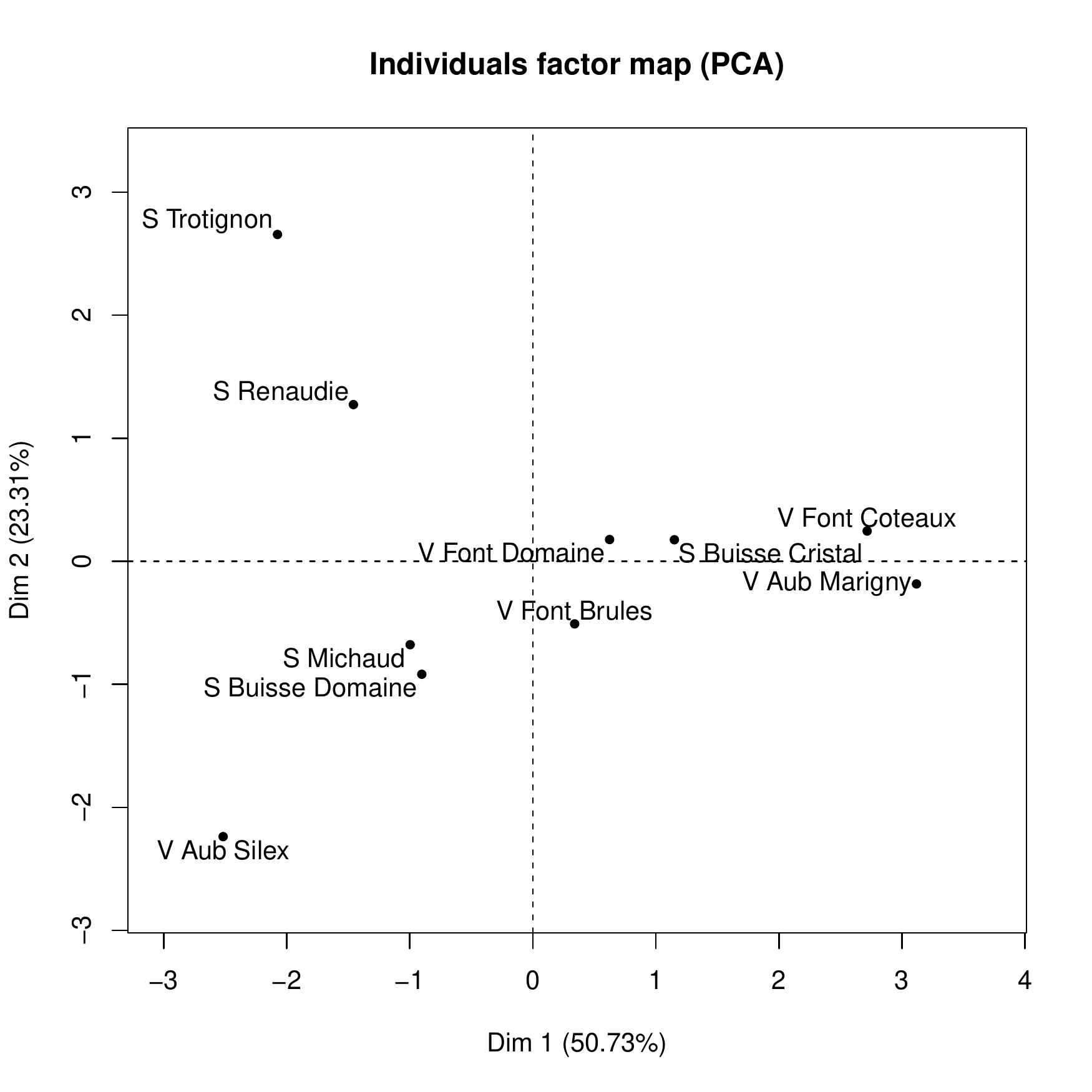}
\caption{2 dimensional wines representation.}
\label{fig:PCAa}
\end{subfigure}
\begin{subfigure}{0.48\textwidth}
\centering
\includegraphics[width=\textwidth]{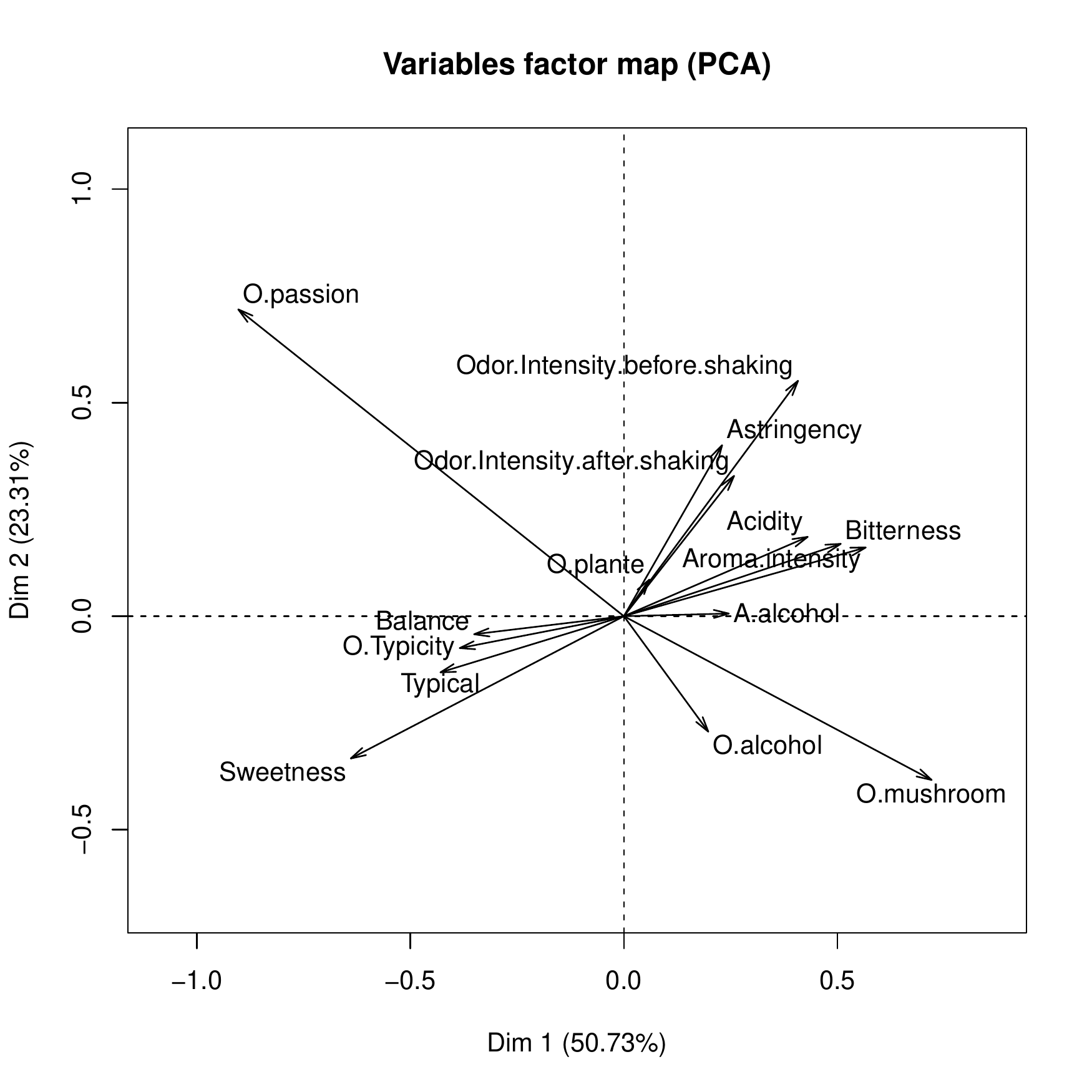}
\caption{PCA variables.}
\end{subfigure}
\caption{PCA on the wine dataset.} \label{fig:pca_consum}
\end{center}
\end{figure}
\spacingset{\SPACEBIG}

We next apply the proposed methods on a dataset from sensory analysis on wines. The data is available from \texttt{http://factominer.free.fr/docs/Consumer\_wine.csv}.
The data was collected by asking consumers to describe 10 white wines (5 Sauvignon and 5 Vouvray) with 15 sensory attributes including sweetness and astringency: each consumer gave each wine a score between 1 and 10 for each attribute. The data was then averaged across consumers into a $10 \times 15$ data matrix $\bfX$, for which we want to perform PCA.
Figure \ref{fig:pca_consum} shows two-dimensional PCA representations, the rows coordinates ($\bfU\boldsymbol{\Lambda^{(1/2)}}$) as well as variables coordinates ($\bfV\boldsymbol{\Lambda^{(1/2)}}$) for this dataset.

Our goal is to associate confidence areas with the low-rank representations shown in Figure \ref{fig:pca_consum}.
Here the 10 wines and 15 attributes are all non-random, so methods based on re-sampling rows of $\bfX$ would not be applicable. Moreover, the effective noise in $\bfX$ gets smaller as the number of interviewed consumers grows; thus the low-noise asymptotic regime studied in Section \ref{sec:asympt} makes sense for this example.

\begin{figure}
\centering
\begin{tabular}{cc}
\includegraphics[width = 0.4\columnwidth]{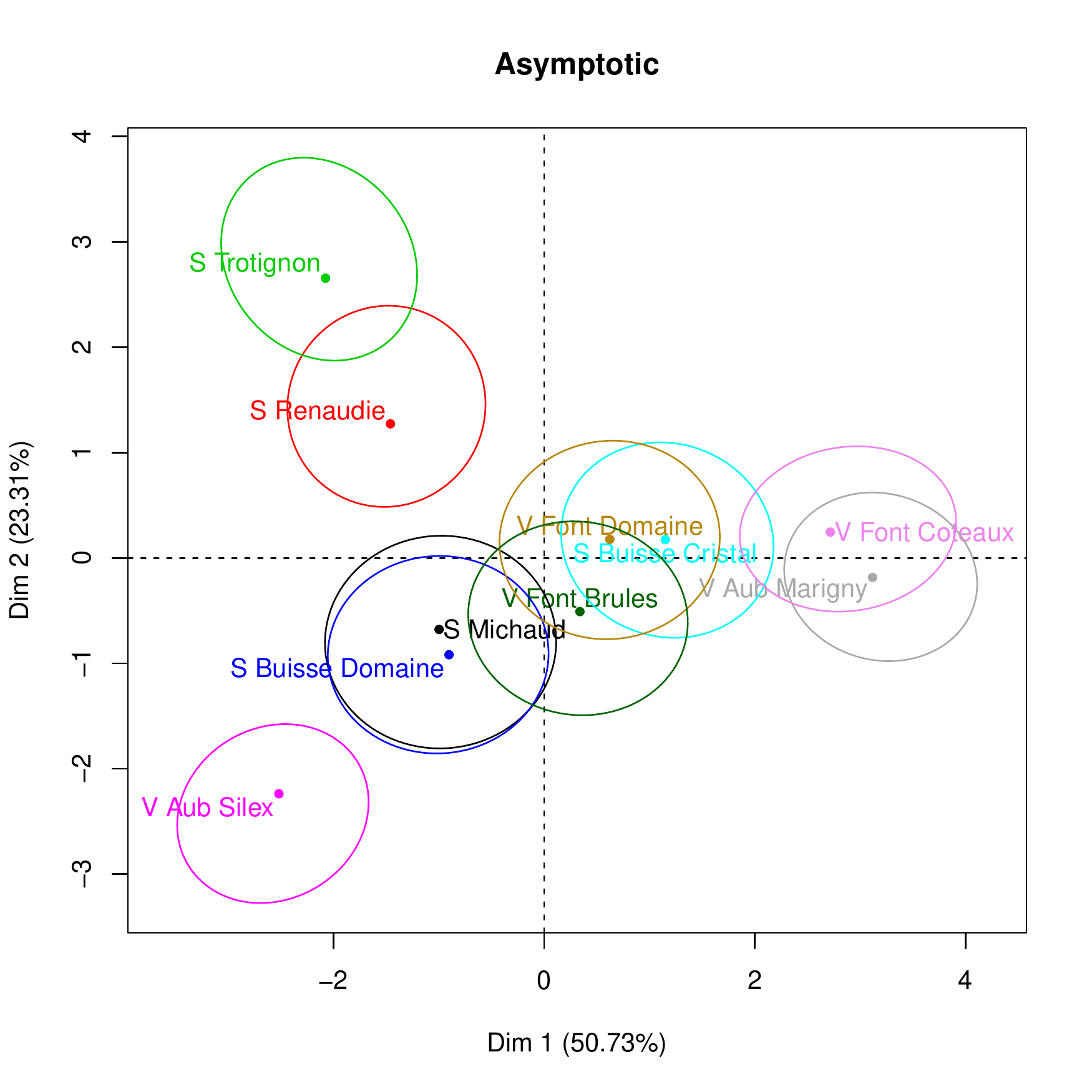} &
\includegraphics[width = 0.4\columnwidth]{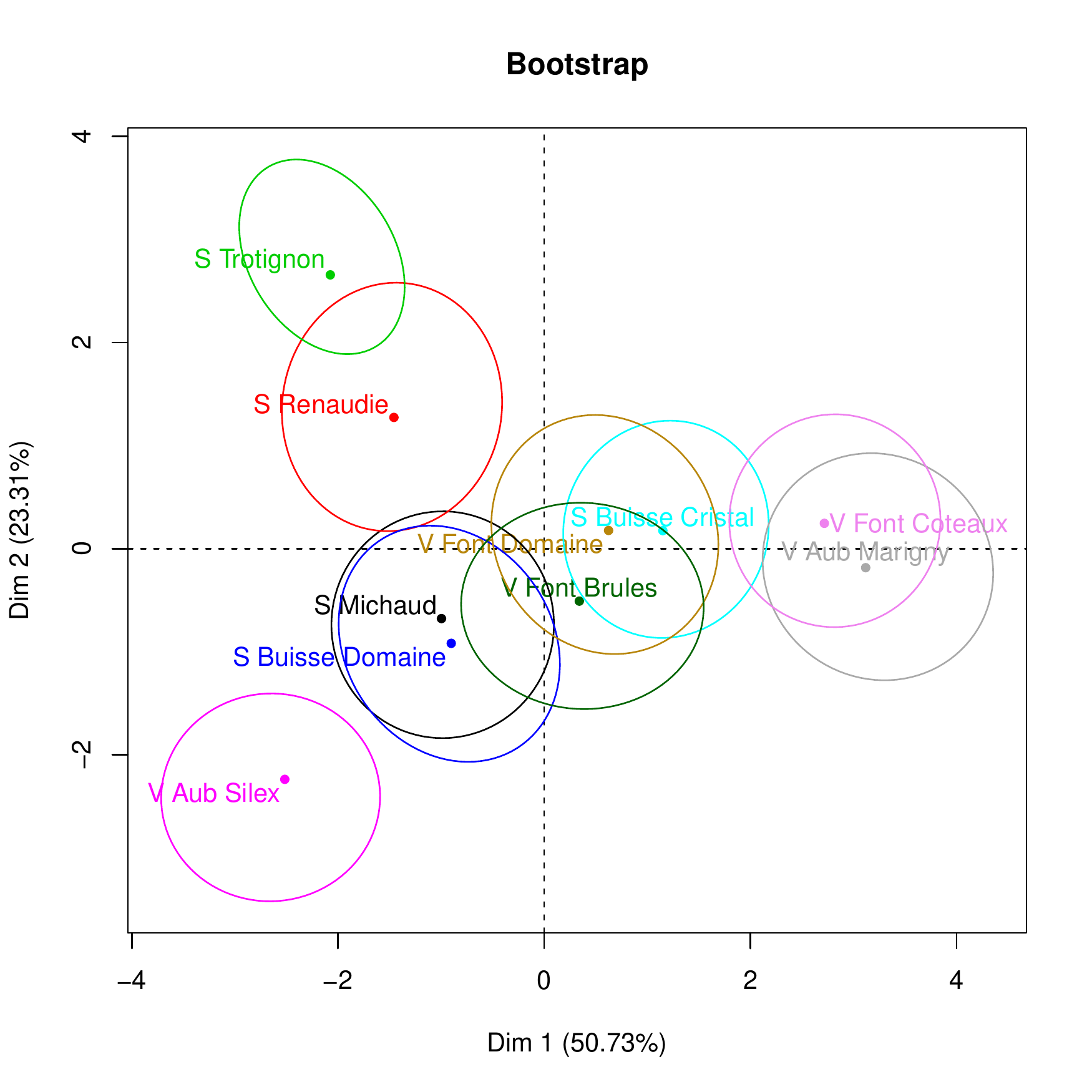} \\
\includegraphics[width = 0.4\columnwidth]{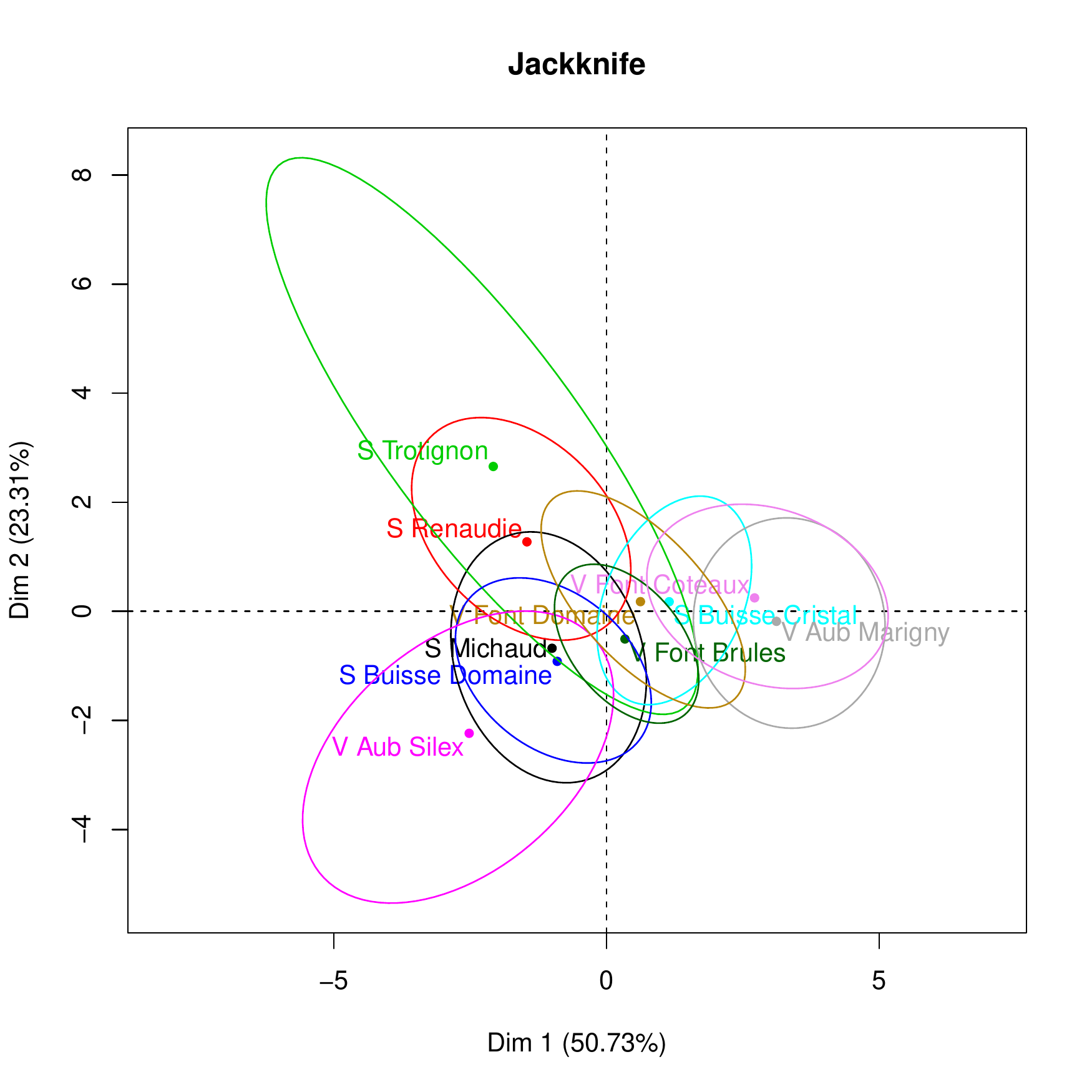} &
\includegraphics[width = 0.4\columnwidth]{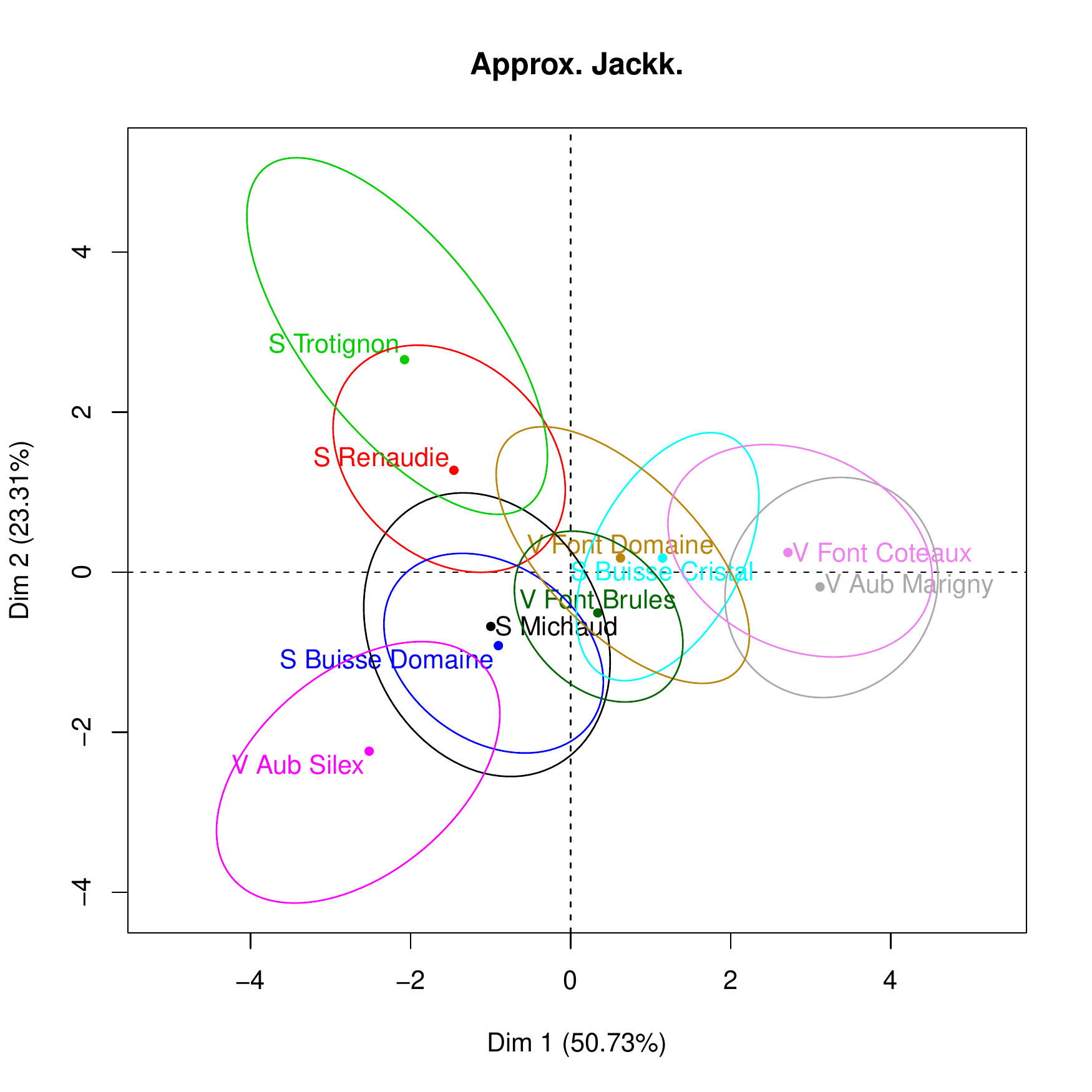} \\
\end{tabular}
\caption{Confidence areas around the wines of the PCA representation using the asymptotic variance, the parametric bootstrap, the jackknife and the approximation of the jackknife.}
\label{fig:areas_wines}
\end{figure}

Figure~\ref{fig:areas_wines} show the result of applying our proposed methods to this dataset. Note that these methods all require us to specify the number of dimensions $S$ as an input parameter. Different methods for doing this are available in the literature \citep{Jolliffe02}; here, we used a cross-validation procedure \citep{Josse11b} which suggested that using $S=2$ dimensions was appropriate.

\spacingset{\SPACESMALL}
\spacingset{\SPACEBIG}

The confidence ellipses produced by the asymptotic method and by the bootstrap are all fairly round. In comparison, the ellipses obtained by the jackknife are larger and are much more elongated. In particular, the jackknife appears to suggest that, for the individuals (Figure~\ref{fig:areas_wines}), there is much less signal in the upper-left direction dominated by the Sauvignon Trotignon and the Sauvignon Renaudie wines than in the other directions. Conversely, all methods seem to imply that the Vouvray Aubence Silex wine in the lower-left corner is perceived differently from most other wines in a significant way, especially from the point of view of the descriptor sweetness. 

\spacingset{\SPACESMALL}
\begin{table}
\begin{center}
\caption{Simulation study, using the rank-two approximation to the wine data matrix as the true signal. The table shows average coverage for the produced confidence ellipsoids; nominal coverage is $1 - \alpha = 0.95$. The true matrix was perturbed with isotropic Gaussian noise of various scales $\sigma$. These results are averaged over 200 replicates.}
\begin{tabular}{r|cccc|}
  \hline
Noise scale $\sigma$ & Asymptotic & Bootstrap & Jackknife & Approx. Jack. \\ 
  \hline
0.1 & \bf 0.942 & \bf 0.938 & 0.974 & \bf 0.95 \\ 
  0.2 & \bf 0.938 & \bf 0.938 & 0.98 & \bf 0.952 \\ 
  0.4 & 0.916 & \bf 0.93 & 0.984 & \bf 0.958 \\ 
  0.6 & \it 0.876 & 0.92 & 0.982 & \bf 0.953 \\ 
  0.8 & \it 0.804 & \it 0.89 & 0.986 & \bf 0.948 \\ 
  1 & \it 0.726 & \it 0.838 & 0.982 & \bf 0.936 \\ 
  1.2 & \it 0.64 & \it 0.816 & 0.986 & \bf 0.934 \\ 
   \hline
\end{tabular}
\label{tab:simu_wine}
\end{center}
\end{table}
\spacingset{\SPACEBIG}

In order to get a better idea of which method was most trustworthy here, we ran a small simulation study built on top of the wine dataset. We used the fitted rank-two PCA matrix as the true signal matrix, and then added isotropic Gaussian noise to it. Results for various noise scales $\sigma$ are shown in Table \ref{tab:simu_wine}.
The root-mean residual for the actual PCA fit to the wine data was $\sigma = 0.6$. Our small simulation seems to suggests that the bootstrap was slightly anti-conservative here, whereas the jackknife was slightly conservative. 
Note the good behavior of the approximate jackknife. 
Thus, the most honest confidence areas for the wine dataset should probably be somewhere in between the bootstrap and jackknife ellipses shown in Figure \ref{fig:areas_wines}. 

\subsection{Decathlon Data}
\label{sec:decathlon}

\begin{figure}
\centering
\begin{tabular}{cc}
\includegraphics[width = 0.5\columnwidth]{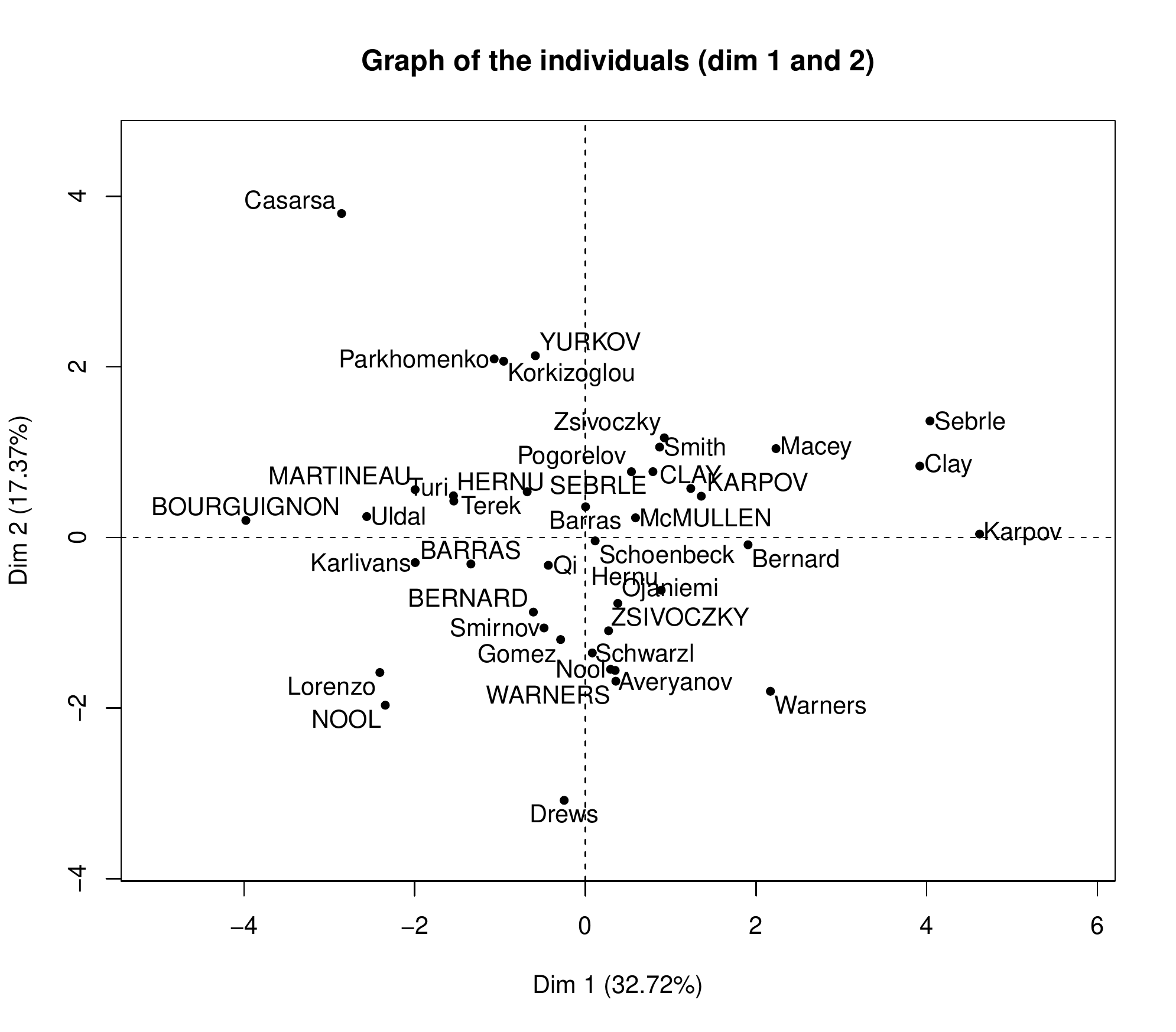} &
\includegraphics[width = 0.43\columnwidth]{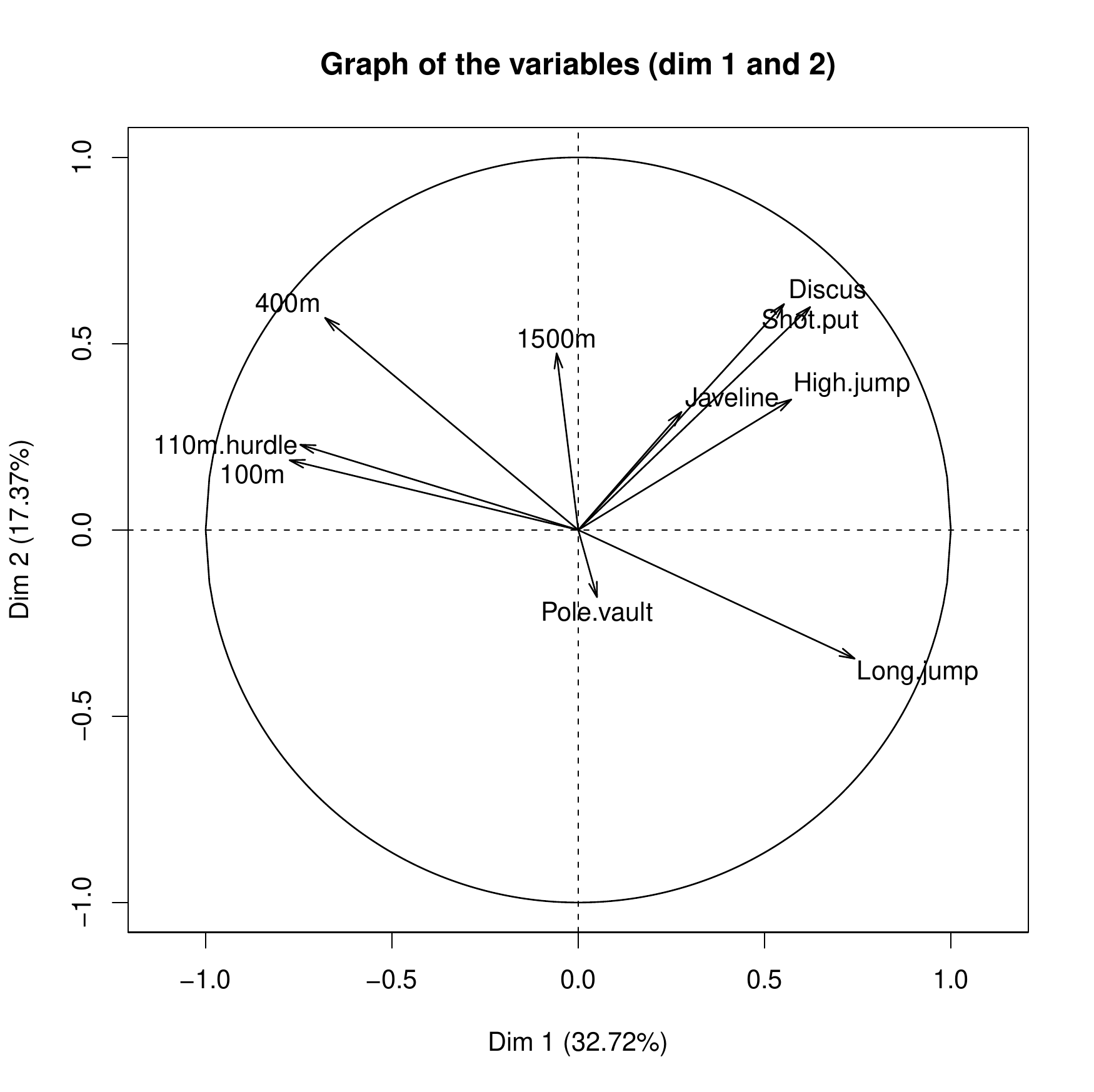} \\
\end{tabular}
\caption{Decathlon dataset: representation of the individuals (left) and of the variables (right) on the first two dimensions of the PCA.}
\label{fig:decat12}
\end{figure}

\begin{figure}
\centering
\begin{tabular}{cc}
\includegraphics[width = 0.5\columnwidth]{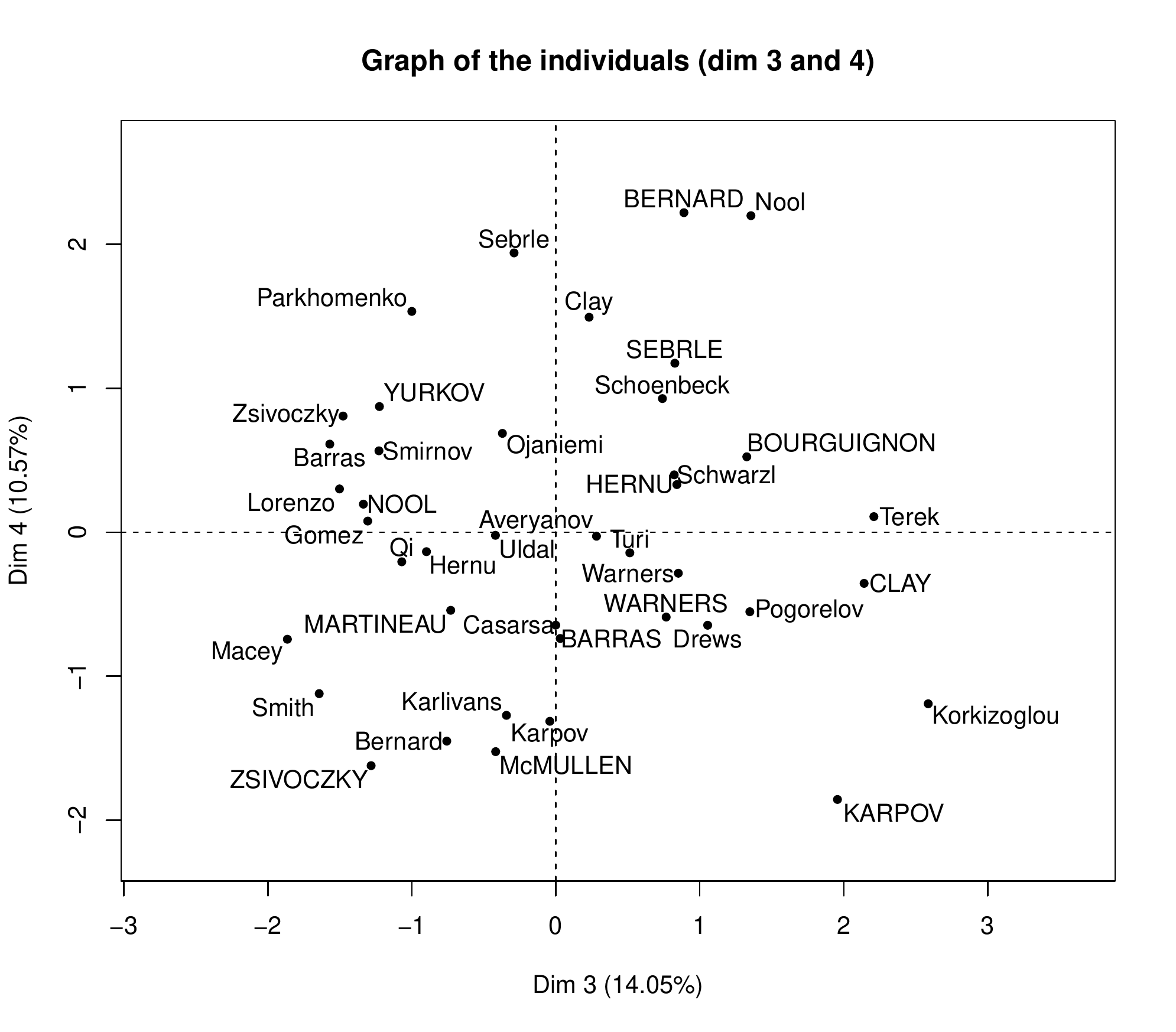} &
\includegraphics[width = 0.43\columnwidth]{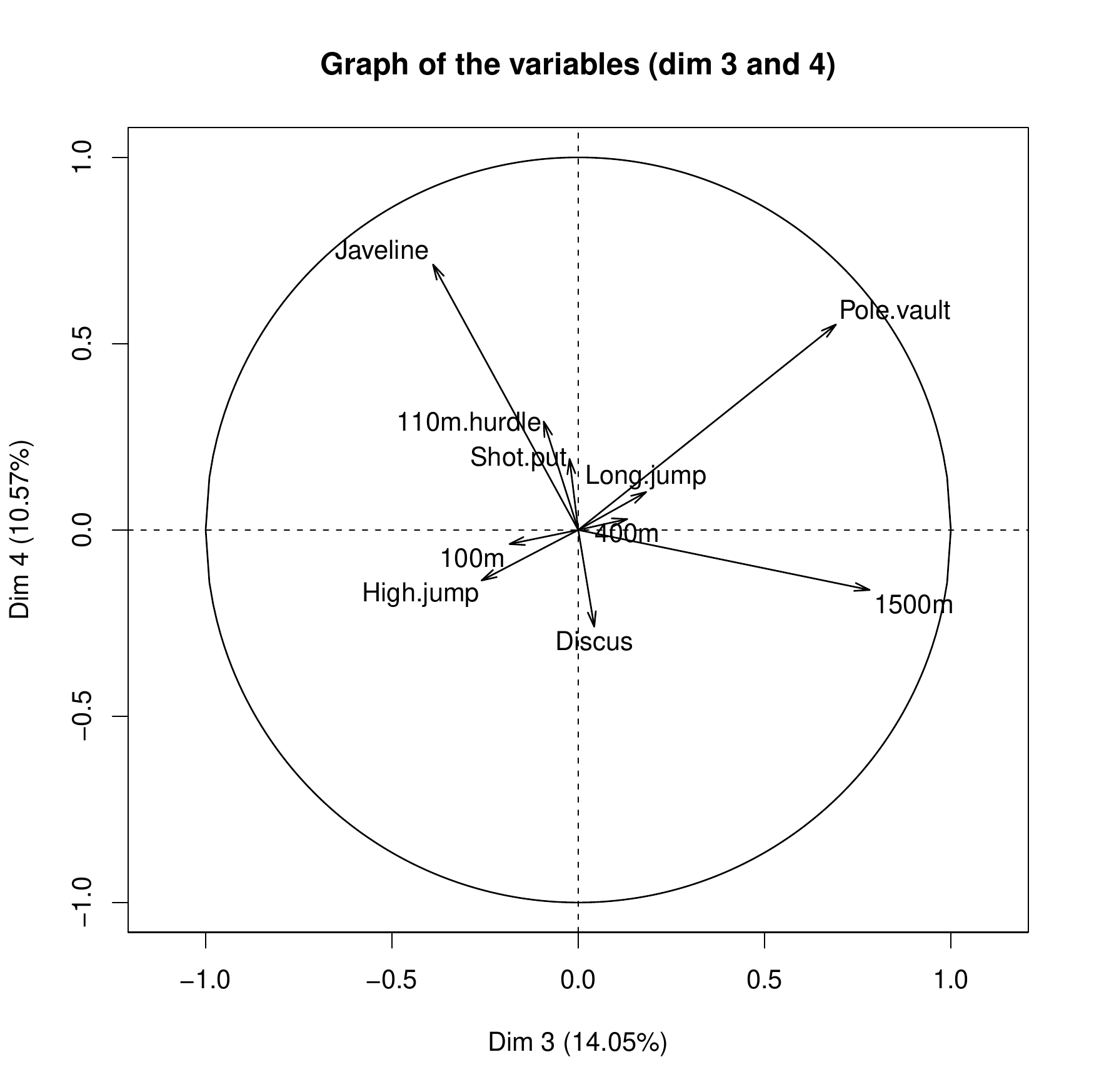} \\
\end{tabular}
\caption{Decathlon dataset: representation of the individuals (left) and of the variables (right) on the dimensions 3 and 4 of the PCA.}
\label{fig:decat34}
\end{figure}

So far, we have focused on the case where the number of dimension $S$ is equal to 2. 
Indeed, when PCA is used as a visualization tool, it is very common to examine and interpret only the first two dimensions.  
This can be explained in part because it is easier from a visualization point of view but also because it makes sense in many applications.  
However some datasets cannot be summarized in two dimensions. 
For instance, let us consider the decathlon dataset (available in the R package FactoMineR \citep{Le:Josse:Husson:2008:JSSOBK:v25i01}) which 
contains the performance of 41 athletes in the different disciplines of the decathlon during the Olympic Game 2004 and the Decastar 2004. The first dimension (see Figure \ref{fig:decat12}) rank the athletes from the best ones on the right who performed well on many trials (they run fast, jump high, threw the discuss far, etc.) to the  ones with lower results on the left. The second dimension opposes the powerful athletes (high values for discuss and shot.put) to the others. The first two dimensions
represent 50\%  of the variability. A greater percentage is not desirable since it would imply that the decathlon does not require different physical aptitudes (if all the trials were highly correlated). The dimensions 3 and 4 (see Figure \ref{fig:decat34}) allow to differentiate athletes who score high at the 1500m and pole vault against those who shine in the javelin trial.  

\spacingset{\SPACESMALL}
\spacingset{\SPACEBIG}

\spacingset{\SPACESMALL}
\spacingset{\SPACEBIG}

The proposed methods (Section \ref{sec:conf}) can be applied straightforwardly to study the variability of the low rank estimator even for a rank greater than 2.  
We illustrate this ability using a small simulation study built on the top of the decathlon dataset in the same way as the simulations for the wine dataset (Section \ref{sec:wine}): an isotropic Gaussian noise is added to the signal in 4 dimensions. Using all the pseudo-realizations, we built ellipsoids in 4 dimensions and we counted the number of times that the row points of the signal are inside the ellipsoids to assess the coverage. The results obtained in Table \ref{tab:simu_deca} are in agreement with the previous results showing results close to the nominal level for all the methods with a slightly underestimation for the bootstrap and an overestimation for the jackknife in high level noise situations. The approximate jackknife shows a good behavior in all the simulations. 

\begin{figure}
\centering
\begin{tabular}{ccc}
\includegraphics[width = 0.25\columnwidth]{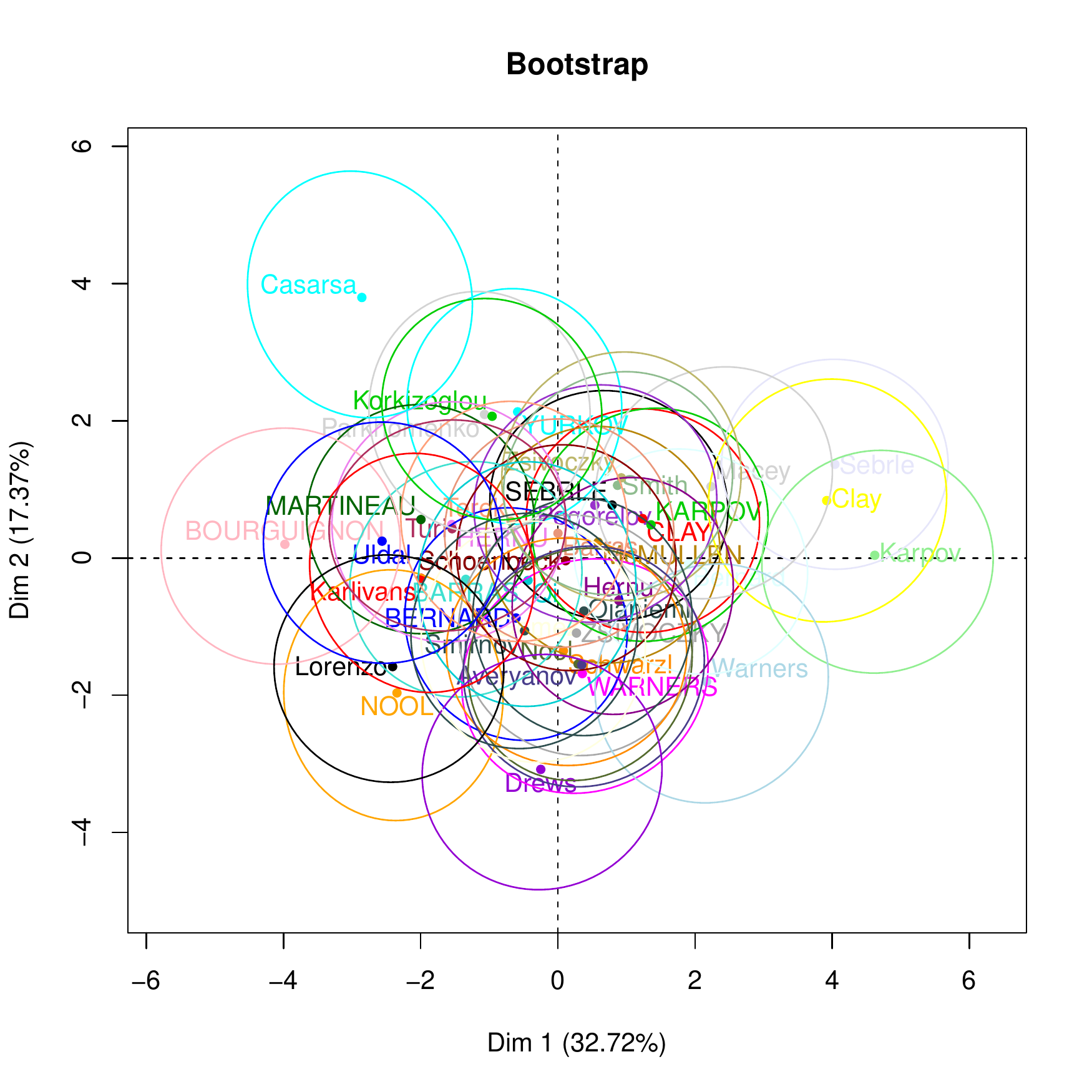} & 
\includegraphics[width = 0.25\columnwidth]{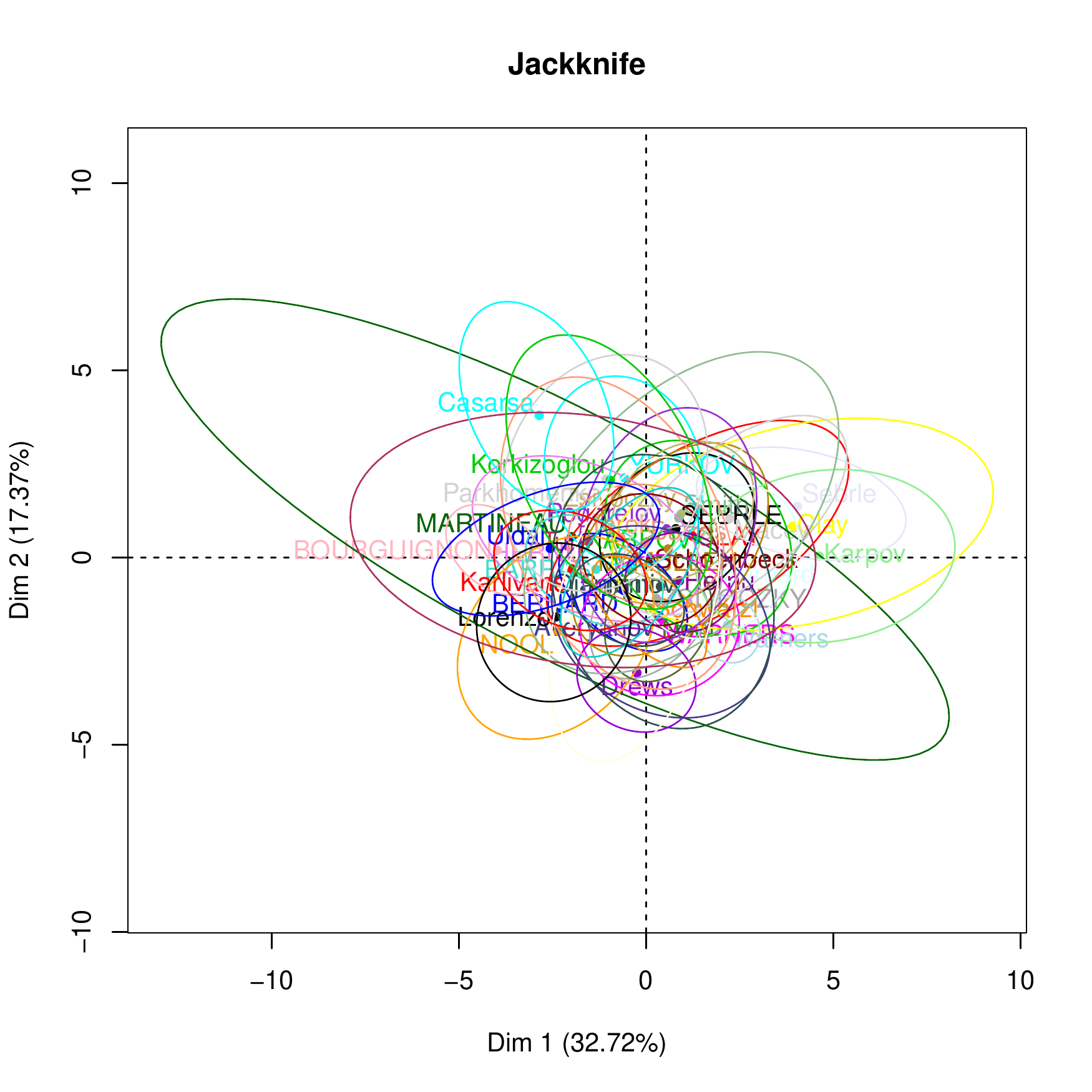} &
\includegraphics[width = 0.25\columnwidth]{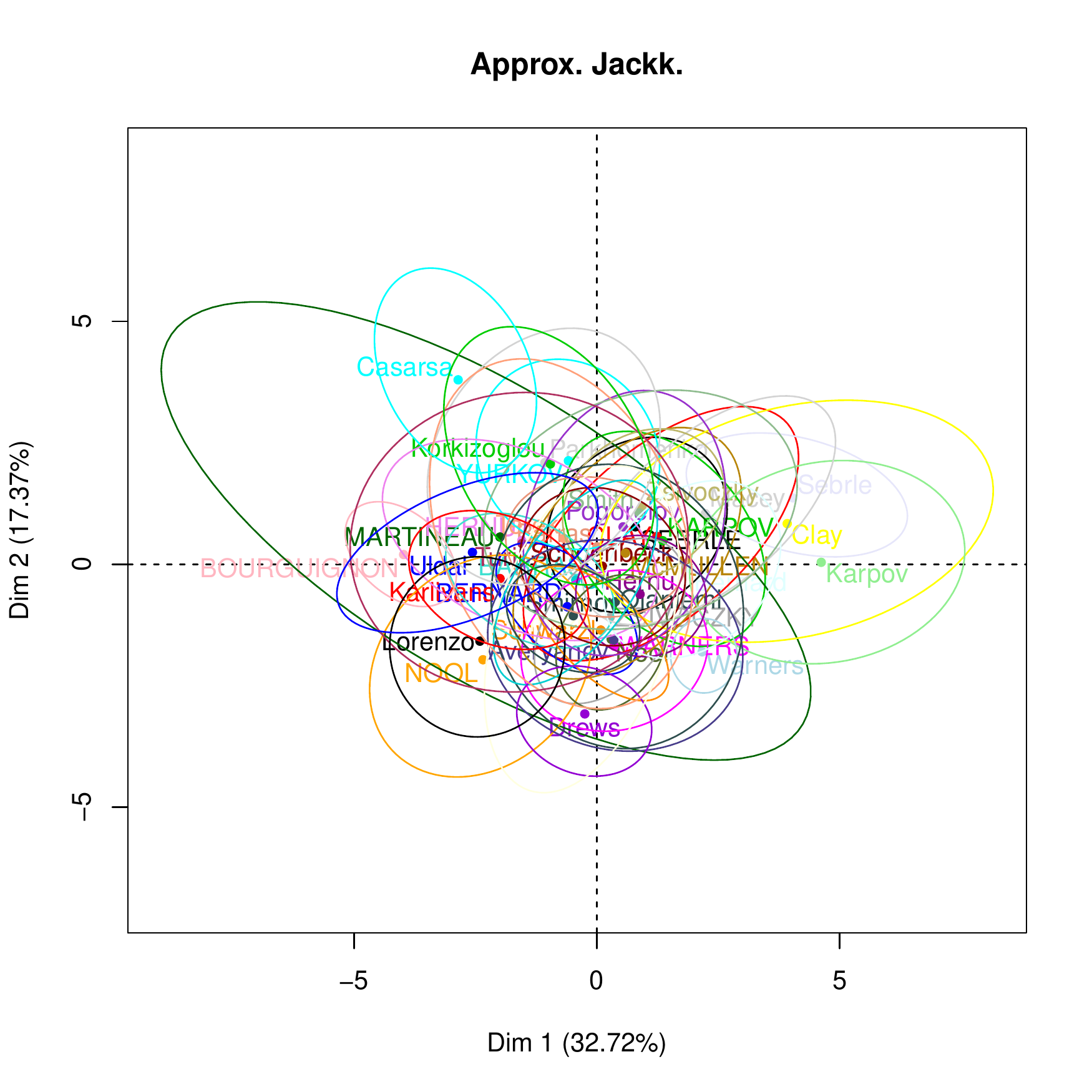} \\
\includegraphics[width = 0.25\columnwidth]{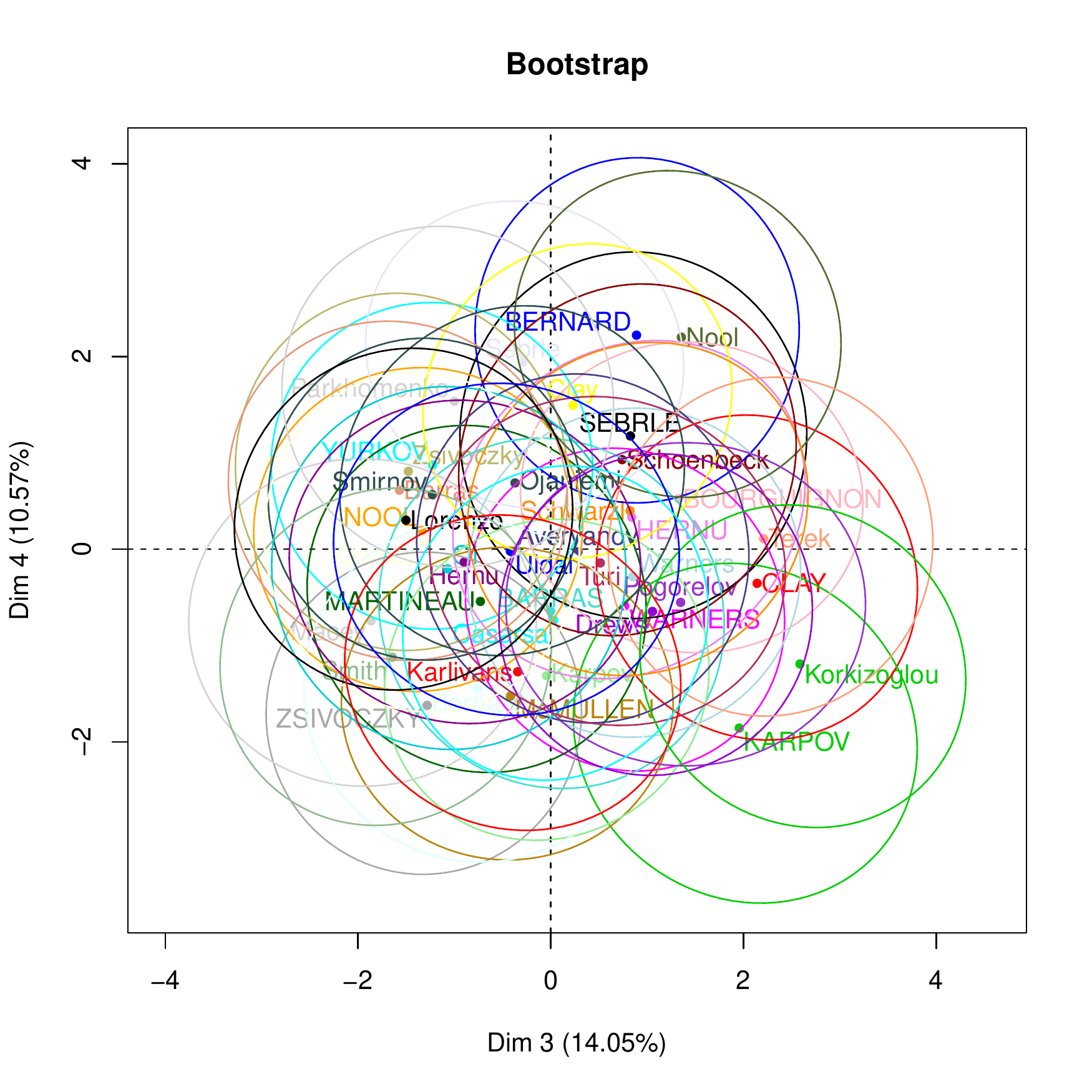} & 
\includegraphics[width = 0.25\columnwidth]{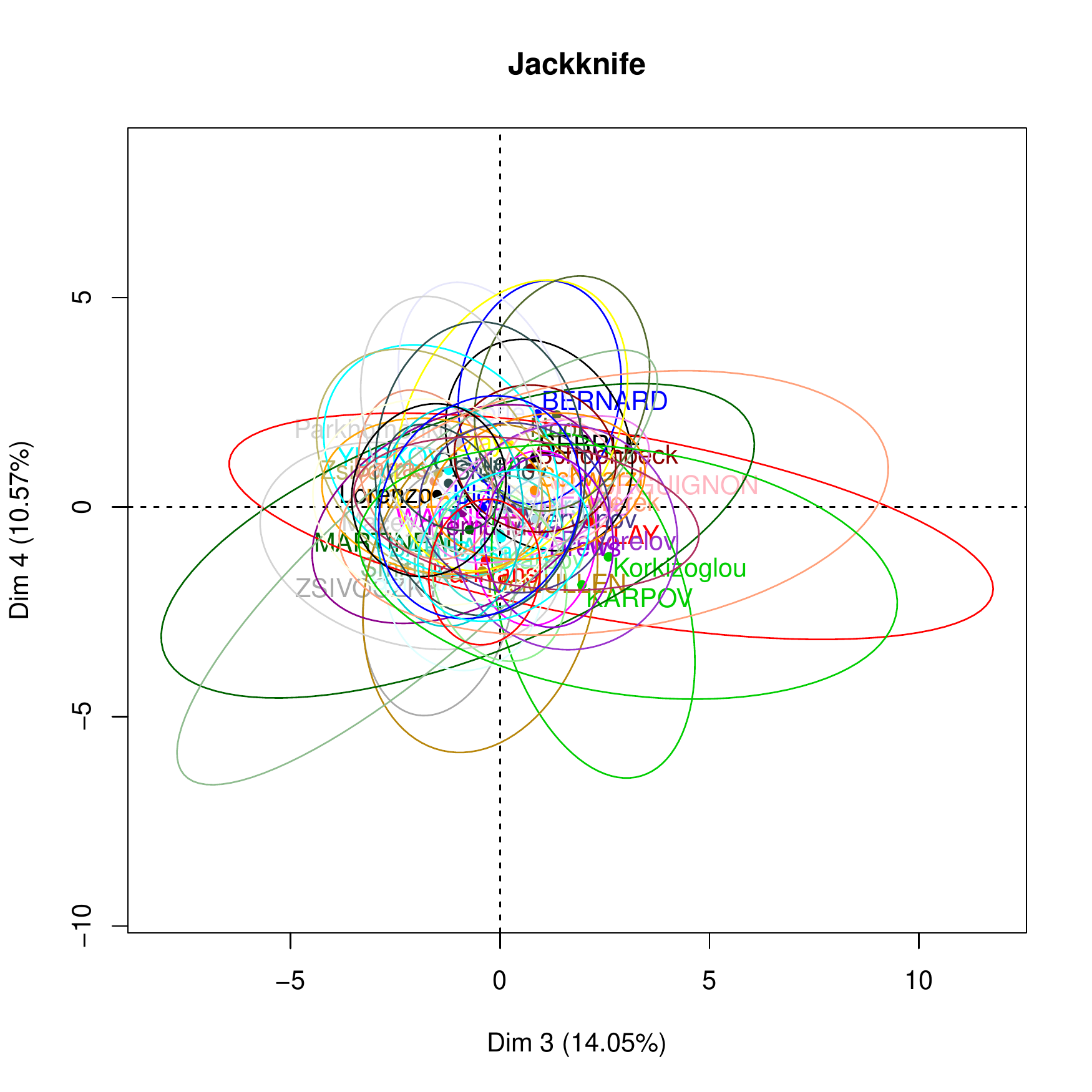} &
\includegraphics[width = 0.25\columnwidth]{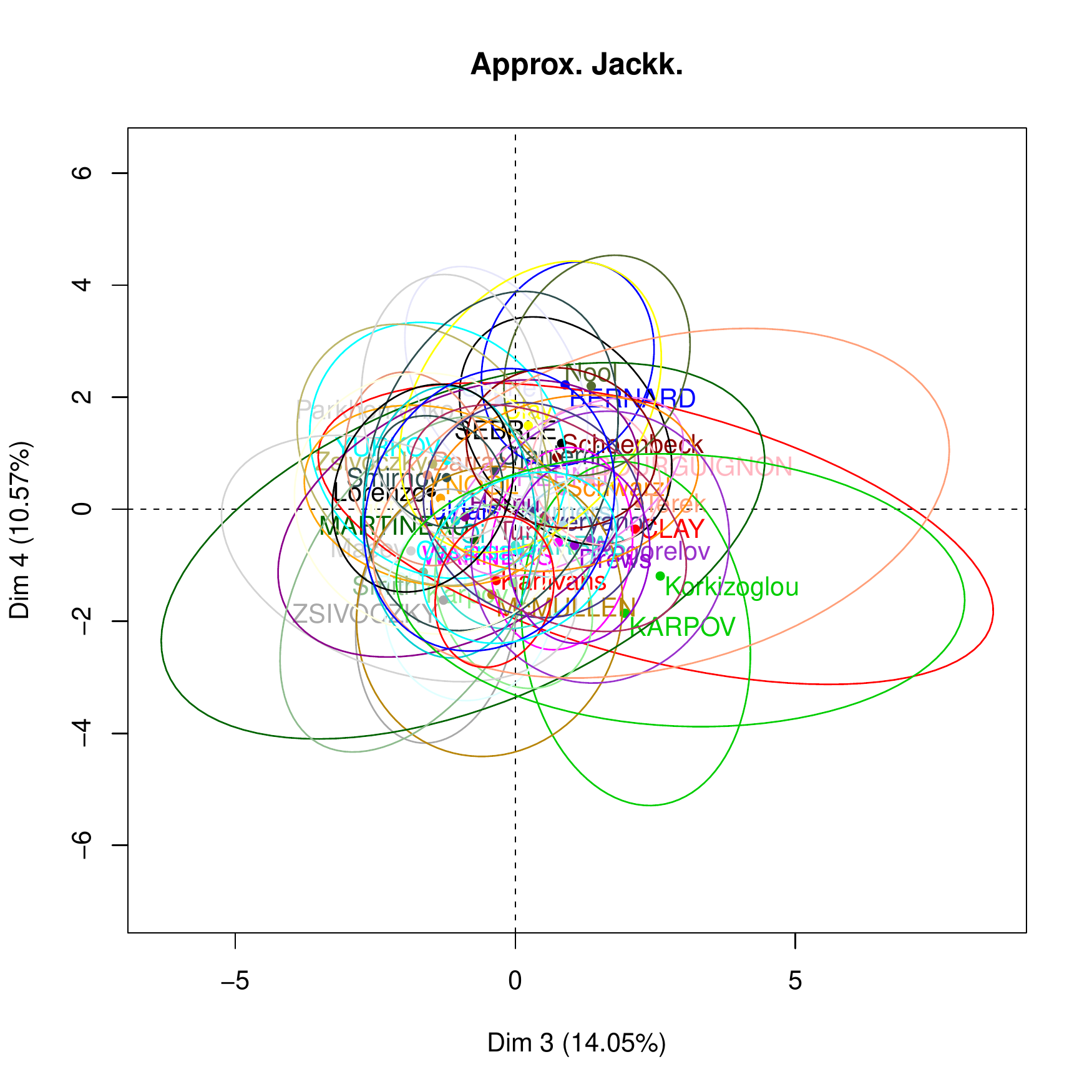} \\

\end{tabular}
\caption{Confidence areas around the athletes of the PCA representation using the parametric bootstrap, the jackknife and the approximation of the jackknife for the dimensions 1-2 on the top and 3-4 on the bottom.}
\label{fig:areas_decat}
\end{figure}

Then we visualize the results on the real dataset representing the ellipsoids around the individuals on the PCA representation on the dimensions 1-2 and on the dimensions 3-4 (Figures \ref{fig:areas_decat}). We do not represent the asymptotic method since the results are very close to the ones obtained by the bootstrap.
Figure~\ref{fig:areas_decat} gives the projections of 4-dimensional ellipsoids in 2 dimensions and thus such areas can only be interpreted as areas of variability and not anymore as confidence areas.  The ellipses obtained by all the methods are quite large suggesting a noisy data set. In such a noisy situation, the simulations (Table \ref{tab:simu_deca}) indicated the bad behavior of the Jackknife with a coverage close to one. This is in agreement with what is observed where some individuals have very large ellipses which leads to graphics that are impossible to interpret and which seem difficult to trust. There are a lot of overlapping between the ellipses on the dimensions 3-4 for the bootstrap and the approximate Jackknife which may suggest that these dimensions are not very significant. Nevertheless, these ellipses allow to differentiate between individuals such as Karpov on the bottom to Nool on the top. 

\spacingset{\SPACESMALL}
\begin{table}
\begin{center}
\caption{Simulation study, using the rank-four approximation to the decathlon data matrix as the true signal. The table shows average coverage for the produced confidence ellipsoids; nominal coverage is $1 - \alpha = 0.95$. The true matrix was perturbed with isotropic Gaussian noise of various scales $\sigma$. These results are averaged over 200 replicates.}
\begin{tabular}{r|cccc|}
  \hline
Noise scale $\sigma$ & Asymptotic & Bootstrap & Jackknife & Approx. Jack. \\ 
  \hline
0.05 & \bf 0.946 & \bf 0.941 & \bf 0.942 & \bf 0.934 \\ 
  0.1 & \bf 0.941 & \bf 0.941 & \bf 0.945 & \bf 0.935 \\ 
  0.2 & \bf 0.936 & \bf 0.939 & \bf 0.949 & \bf 0.941 \\ 
  0.3 & \bf 0.937 & \bf 0.939 & \bf 0.952 & \bf 0.942 \\ 
  0.4 & \bf 0.932 & \bf 0.938 & \bf 0.957 & \bf 0.941 \\ 
  0.6 & 0.91 & 0.921 & \bf 0.963 & \bf 0.939 \\ 
  0.8 & \it 0.897 & 0.912 & 0.973 & \bf 0.946 \\ 
  1 & \it 0.86 & \it 0.891 & 0.98 & \bf 0.935 \\ 
  1.2 & \it 0.827 & \it 0.862 & 0.982 & \bf 0.931 \\ 
   \hline
\end{tabular}
\label{tab:simu_deca}
\end{center}
\end{table}
\spacingset{\SPACEBIG}

\spacingset{\SPACESMALL}
\spacingset{\SPACEBIG}

\section{Discussion}

In this paper, we introduced and studied various approaches to estimating the variance of the PCA estimator and suggested a solution for visualizing these estimates as confidence ellipsoids on PCA maps, emphasizing that confidence areas ensure relevant interpretation. We compared coverage properties of confidence areas based on an asymptotic method, a parametric bootstrap, a jackknife and an approximation to the jackknife, and found that the resulting variance estimates vary widely. 
The asymptotic method and the bootstrap do well in low-noise setting, but can fail when the noise level gets high or when the number of variables is much greater than the number of rows. On the other hand, the jackknife has good coverage properties for large noisy examples but requires a minimum number of variables to be stable enough.

All our methods rely on the validity of the model  \eqref{mod_acp}, and consequently the choice of the number of dimension $S$ is crucial. We focused mainly on the case where $S=2$ on the simulations. 
However, we showed that the methods can still be applied and are accurate for a number greater than 2. 
The results are then visualized with areas of variability.

Note that the principal components  of  $\bfX$ obtained by PCA are unbiased estimates of the true principal components of $\tilde\bfX$ in the asymptotic regime when the noise variance tends to 0.
But when the signal to noise ratio is small, the principal components of $\bfX$ are biased and consequently the confidence regions centered on these principal components are not necessarily optimal. Nevertheless, even when they can no longer be considered as confidence regions for the true underlying principal components, they can still be used to visualize the variance of the PCA estimator due to noise.

There appear to be many potential applications for our methods. For example, in the field of plant breeding, biplots are often used to represent the interaction between genotypes and environments; in this case, having access to confidence areas on the biplot representation is important for protection against spurious interpretations about, say, differences between genotypes.
Other approaches to this problem include the work of \citet{Perez11} and \citet{Josse14} who suggested a Bayesian treatment of the model enabling them to get direct distributions for the parameters, and that of \citet{Yang13} who proposed another bootstrap approach which consists in simultaneously bootstrapping rows and columns of the data matrix.

We finish by discussing some opportunities for further research. Extending our method to account for uncertainty in the number of dimensions $S$ presents interesting challenges. One idea would be be to build on ideas of \citet{efron14}, who discusses the estimation of prediction error when the model that was fit is itself a random variable. Another solution would be to use a fully Bayesian approach in the line with the works of  \citet{Hoff2007, hoff2009} for models based on singular-value decompositions. Visualizing confidence areas when $S$ is random may also prove to be difficult

Second, our methods could be used for studying the variability of other estimators related to PCA. For instance, several estimators that apply different thresholding and/or shrinking rules on the empirical singular values while keeping the empirical singular vectors unchanged have been proposed for denoising data in the context of low-rank matrix estimation \citep{Mazumder:STMiss:2010,Verbanck:RegPCA:2013, Donoho:SVDHT:2013, Sourav:2013, JosseSardy:GSURE:2013, RajRao:2013,  Shabalin2013}. \citet{Shabalin2013} proposed  an estimator that explicitly accounts for the effect of the noise on the singular values and on the singular vectors.  This family of estimators can provide a better recovery of the underlying signal in term of mean squared error in comparison to PCA. These estimators reduce variance at the price of bias. Studying the variability of these estimators with the proposed approaches as well as their coverage properties could lead to useful insights.

Finally, we note that it may be worth using the values of the projection matrix  $\bfP_{ij,ij}$ as leverage values.  Of course, these can only be seen as approximate measures of leverage since they are based on a linear approximation to the expectation surface. Nevertheless, they can be used to define corrected residuals $\hat \varepsilon_{ij}=\frac{\hat x_{ij}-x_{ij}}{\sqrt{1-\bfP_{ij,ij}}}$ or Cook's distances in the framework of PCA. In addition, in non-linear regression, such values are related to curvature measures \citep{Lau93}, and can be used to assess the impact of leverage points and outliers on the jackknife \citep{Simonoff86}.
More research can be done in these directions.

\section*{Acknowledgements}
Julie Josse has received the support of the European Union, in the framework of the Marie-Curie FP7 COFUND People Programme, through the award of an AgreenSkills' fellowship (under grant agreement n° 267196) for an academic
visit to Stanford. 

\bibliographystyle{spbasic}  
\bibliography{josse} 

\end{document}

%% file: table1.tex
\begin{table}[p]
\caption{Simulated coverage for the asymptotic method, the bootstrap, the jackknife, and the approximation of the jackknife. Nominal coverage is $1 - \alpha = 0.95$, and results are averaged over 50 replicates. We vary the signal-to-noise ratio (SNR), the relative scale of the first and second principal components ($d_1/d_2$), and the number of variables $p$ and rows $n$. In bold, values greater than 0.93 and less than 0.97;
in italic values less than 0.90. All simulations use a two-dimensional Gaussian signal as the true signal. In some cases, the exact jackknife was too slow so we only ran its approximation.}
\label{tab:coverage}
\begin{center}
\begin{tabular}{||rr|rr||cc|cc||}
SNR & d1/d2 & p & n & Asymp & Boot & Jack & Approx Jack \\ 
  \hline \hline
1.00 & 1.00 &   5 &  20 & \it 0.889 & 0.92 & 0.919 & \it 0.863 \\ 
  1.00 & 1.00 &   5 &  50 & 0.916 & 0.928 & \it 0.836 & \it 0.817 \\ 
  1.00 & 1.00 &   5 & 100 & \bf 0.931 & \bf 0.935 & \it 0.797 & \it 0.788 \\ 
  1.00 & 1.00 &  20 &  20 & \it 0.851 & 0.902 & \bf 0.948 & 0.924 \\ 
  1.00 & 1.00 &  20 &  50 & 0.917 & \bf 0.935 & 0.925 & 0.916 \\ 
  1.00 & 1.00 &  20 & 100 & \bf 0.93 & \bf 0.94 & - & 0.917 \\ 
  1.00 & 1.00 & 100 &  20 & \it 0.786 & \it 0.852 & - & 0.97 \\ 
   \hline
1.00 & 4.00 &   5 &  20 & \it 0.856 & \it 0.892 & \bf 0.962 & \it 0.896 \\ 
  1.00 & 4.00 &   5 &  50 & 0.913 & 0.927 & \bf 0.959 & 0.914 \\ 
  1.00 & 4.00 &   5 & 100 & \bf 0.931 & \bf 0.937 & \bf 0.956 & 0.923 \\ 
  1.00 & 4.00 &  20 &  20 & \it 0.81 & \it 0.892 & \bf 0.957 & \bf 0.931 \\ 
  1.00 & 4.00 &  20 &  50 & 0.901 & \bf 0.931 & 0.919 & 0.905 \\ 
  1.00 & 4.00 &  20 & 100 & 0.924 & \bf 0.936 & - & \it 0.896 \\ 
  1.00 & 4.00 & 100 &  20 & \it 0.716 & \it 0.835 & - & 0.976 \\ 
   \hline
\hline
4.00 & 1.00 &   5 &  20 & \bf 0.938 & \bf 0.938 & \it 0.868 & \it 0.849 \\ 
  4.00 & 1.00 &   5 &  50 & \bf 0.935 & \bf 0.935 & \it 0.805 & \it 0.796 \\ 
  4.00 & 1.00 &   5 & 100 & \bf 0.94 & \bf 0.941 & \it 0.769 & \it 0.765 \\ 
  4.00 & 1.00 &  20 &  20 & \bf 0.938 & \bf 0.941 & \bf 0.951 & \bf 0.933 \\ 
  4.00 & 1.00 &  20 &  50 & \bf 0.943 & \bf 0.944 & \bf 0.933 & 0.927 \\ 
  4.00 & 1.00 &  20 & 100 & \bf 0.942 & \bf 0.945 & - & 0.925 \\ 
  4.00 & 1.00 & 100 &  20 & \bf 0.935 & \bf 0.936 & - & \bf 0.953 \\ 
   \hline
4.00 & 4.00 &   5 &  20 & 0.926 & 0.924 & 1 & 0.999 \\ 
  4.00 & 4.00 &   5 &  50 & \bf 0.943 & \bf 0.941 & 0.998 & 0.996 \\ 
  4.00 & 4.00 &   5 & 100 & \bf 0.938 & \bf 0.937 & 0.999 & 0.998 \\ 
  4.00 & 4.00 &  20 &  20 & \bf 0.934 & \bf 0.938 & \bf 0.945 & 0.929 \\ 
  4.00 & 4.00 &  20 &  50 & \bf 0.938 & \bf 0.941 & 0.91 & 0.902 \\ 
  4.00 & 4.00 &  20 & 100 & \bf 0.944 & \bf 0.943 & - & 0.9 \\ 
  4.00 & 4.00 & 100 &  20 & 0.925 & \bf 0.932 & - & \bf 0.957 \\ 
   \hline \hline
\end{tabular}
\end{center}
\end{table}